\documentclass[epj]{svjour}
\usepackage{graphicx}
\usepackage{natbib}
\usepackage{color}
\usepackage{amssymb}
\usepackage{amsmath}
\begin{document}
\title{Diffraction from the $\beta$-sheet crystallites in spider silk}

\author{S.~Ulrich \inst{1} \and A.~Gli\v{s}ovi\'c \inst{2} \and T.~Salditt \inst{2} \and A.~Zippelius \inst{1,3}
}                     
%
%
\institute{Institut f\"{u}r Theoretische Physik \and Institut f\"{u}r R\"{o}ntgenphysik \\Georg-August-Universit\"{a}t G\"{o}ttingen, \\Friedrich-Hund-Platz 1, 37077 G\"{o}ttingen, Germany \and Max-Planck-Institut f\"ur Dynamik und Selbstorganisation \\Bunsenstrasse 10, 37073 G\"ottingen, Germany}
\date{Received: 18 March 2008 / Published online: 10 October 2008}
%
\abstract{
  We analyze the wide angle x-ray scattering from oriented spider silk
  fibers in terms of a quantitative scattering model, including both
  structural and statistical parameters of the $\beta$-sheet
  crystallites of spider silk in the amorphous matrix.  The model is
  based on kinematic scattering theory and allows for rather general
  correlations of the positional and orientational degrees of freedom,
  including the crystallite's size, composition and dimension of the
  unit cell. The model is evaluated numerically and compared to
  experimental scattering intensities allowing us to extract the
  geometric and statistical parameters. We show explicitly that for the
  experimentally found mosaicity (width of the orientational
  distribution) inter-crystallite effects are negligible and the data can
  be analyzed in terms of single crystallite scattering, as
  is usually assumed in the literature.
} 
\maketitle
\section{Introduction}

Spider silk is a material which is since long known to everybody, but
which only more recently receives great appreciation by
the scientific community for its outstanding material properties
\cite{grubb97,properties}.  Interest here has focused on the so-called
\emph{dragline} fiber, the high strength fibers which orb web spiders produce
from essentially only two proteins to build their net's frame and
radii, and also to support their own body weight after an intentional
fall down during escape.  Evolution has optimized dragline fibers for
tensile strength, extensibility and energy dissipation. Dragline silk
can support relatively large strains and has a tensile strength
comparable to steel or Kevlar. For the energy density which can be
dissipated in the material before breaking, the so-called tenacy
(toughness), values of $160\,{\rm MJ/m^3}$ have been reported
\cite{gosline,silkpolymers}, e.g.\@ for different \emph{Nephila}
species, on which most studies have been carried out.  An
understanding of the structural origins of these mechanical properties
is of fundamental interest, and may at the same time serve the
development of biomimetic material design
\cite{scheibel2004,scheibel2006}, using recombinant and synthetic
approaches \cite{ScheibelAPA,KaplanAPA,BauschAPA}.  As for other
biomaterials, the correlation between structure and the mechanical
properties can only be clarified by advanced structural
characterization accompanied by numerical modelling. To this end, not
only the mechanical properties \cite{VollrathAPA,ZbilutAPA} resulting
from the structure, but also the structure itself has to be
modelled to exploit and to interpret the experimental data.  Such
efforts have in the past led to a quantitative understanding of many
biomaterials like bone, tendons and wood \cite{fratzl1,fratzl2}.

As deduced from x-ray scattering
\cite{warwicker,grubb97,becker,grubb_jelinski1} and NMR experiments
\cite{nmr}, spider silks are characterized by a seemingly rather
simple design: the alanine-rich segments of the fibroin polypeptide
chain fold into $\beta$-sheet nano-crystallites (similar to
poly-L-alanine crystals) embedded in an amorphous network of chains
(containing predominately glycine). The crystalline component makes up
an estimated 20\%-30\% of the total volume, and may represent
cross-links in the polymer network, interconnecting several different
chains.  At the same time the detailed investigation of the structure
is complicated, at least on the single fiber level, by the relatively
small diameters in the range of $1-10\,{\rm \mu m}$, depending on the
species.  Using highly brilliant microfocused synchrotron radiation,
diffraction patterns can be obtained not only on thick samples of
fiber bundles, but also on a single fiber
\cite{vollrath1999,vollrath2000,vollrath2001,riekel99,vollrath2004,vollrath2005}.
Single fiber diffraction was then used under simultaneous controlled
mechanical load in order to investigate changes of the molecular
structure with increasing strain up to failure \cite{Glisovic08}.
Note that single fiber diffraction, where possible, is much better
suited to correlate the structure to controlled mechanical load, since
the strain distribution in bundles is intrinsically inhomogeneous, and
the majority of load may be taken up by a small minority of fibers.

While progress of the experimental diffraction studies has been
evident, the analysis of the data still relies on the classical
classification and indexing scheme introduced by
Warwicker. According to Warwicker, the $\beta$-sheet cristallites of the dragline of Nephila
fall into the so-called system 3 of a nearly orthorhombic unit cell
\cite{warwicker, marsh} with lattice constants $10.6\times
9.44 \times 6.95$ \AA\,\cite{warwicker}.  To fix the coordinate
system, they define the $x$-axis to be in the direction of the amino
acid side chains connecting different $\beta$-sheets with a lattice
constant of $a_x=10.6$ \AA\, while the $y$-axis denotes the direction
along the hydrogen bonds of the $\beta$-sheets with a lattice constant
of $a_y=9.44$ \AA. Finally the $z$-axis corresponds to the axis along
the covalent peptide bonds (main chain) with a lattice constant
$a_z=6.95$ \AA. The $z$-axis with small lattice constant is
well-aligned along the fiber axis. Note that while we follow this commen 
convention, other notations and choices of axis are also used in the
literature.  While helpful, the indexing scheme does not give any
precise information on the exact structure of the unit cell, and on
the fact whether the $\beta$-pleated sheets are composed of parallel
or antiparallel strands, and how the two-dimensional sheets are
arranged to stacks. To this end, not only peak positions but the
entire rather broad intensity distribution has to be analyzed.  To
interpret the scattering image it is essential to know, whether
correlations between different crystallites are important or whether
the measured data can be accounted for by the scattering of single
crystallites, averaged over fluctuating orientations.  It is also not
clear, whether correlations between translational and rotational degrees
of freedom are important. Finally, the powder averaging taking into
account the fiber symmetry experimental mosaicity (orientational
distribution) must be quantitatively taken into account.

In this work we built a scattering model based on kinematic scattering
theory and compare the numerically calculated scattering intensity
with the experimental wide angle scattering distribution measured from
aligned silk fibers. The numerical calculations allow for a
quantitative comparison to the experimental data and yield both
structural and statistical parameters.  Note that the small size of
crystallites, leading to correspondingly broad reflections, and a
generally rather low number of external peaks exclude a standard
crystallographic approach. The structural parameters concern the
crystal structure, in particular the atomic positions in the unit
cell, and the crystallite size. The composition is assumed to be that
of ideal polyalanine without lattice defects.  This assumption is
partly justified by the fact that the small size of the crystallite is
the 'dominating defect' in this material.  The statistical parameters
relate to the orientational distribution of the crytallite symmetry
axis with respect to the fiber axis and the correlations between
crystallites. The model is constructed based on a quite general
approach, allowing independently for correlations between center-of-mass
positions (translations) and crystallite orientations (rotations). 

The paper is organized as follows. In Sec.~\ref{sec:theory} we
introduce the basic model with parameters for the crystallite size,
lattice constants and statistical parameters for the crystallites'
position and orientation. Subsequently, in Sec.~\ref{sec:sf} we
compute the scattering function for our model. Sec.~\ref{sec:ConfigurationUnitCell} specifies the different atomic
configurations which are conceivable for polyalanine. The main results
and the comparison of calculated and measured intensities are
presented in section \ref{sec:results}, before the paper closes with a
short conclusion.

\label{sec:introduction}
\section{Model}
\label{sec:theory}

\begin{figure}[h]
 \begin{minipage}[t]{.45\textwidth}
   \includegraphics[width=3cm]{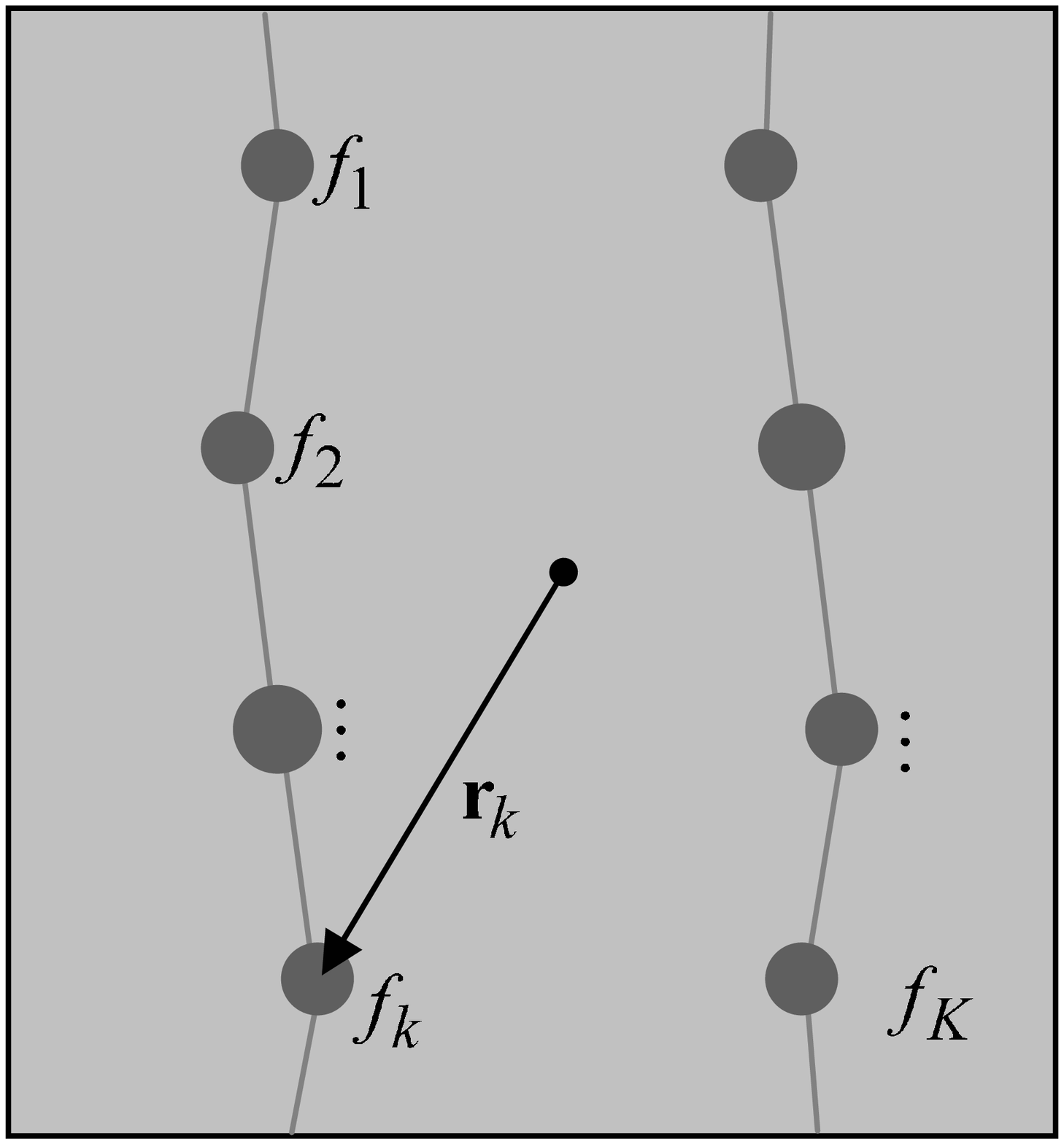}
   \centering
 \end{minipage}

 \begin{minipage}[t]{.45\textwidth}
   \includegraphics[width=5cm]{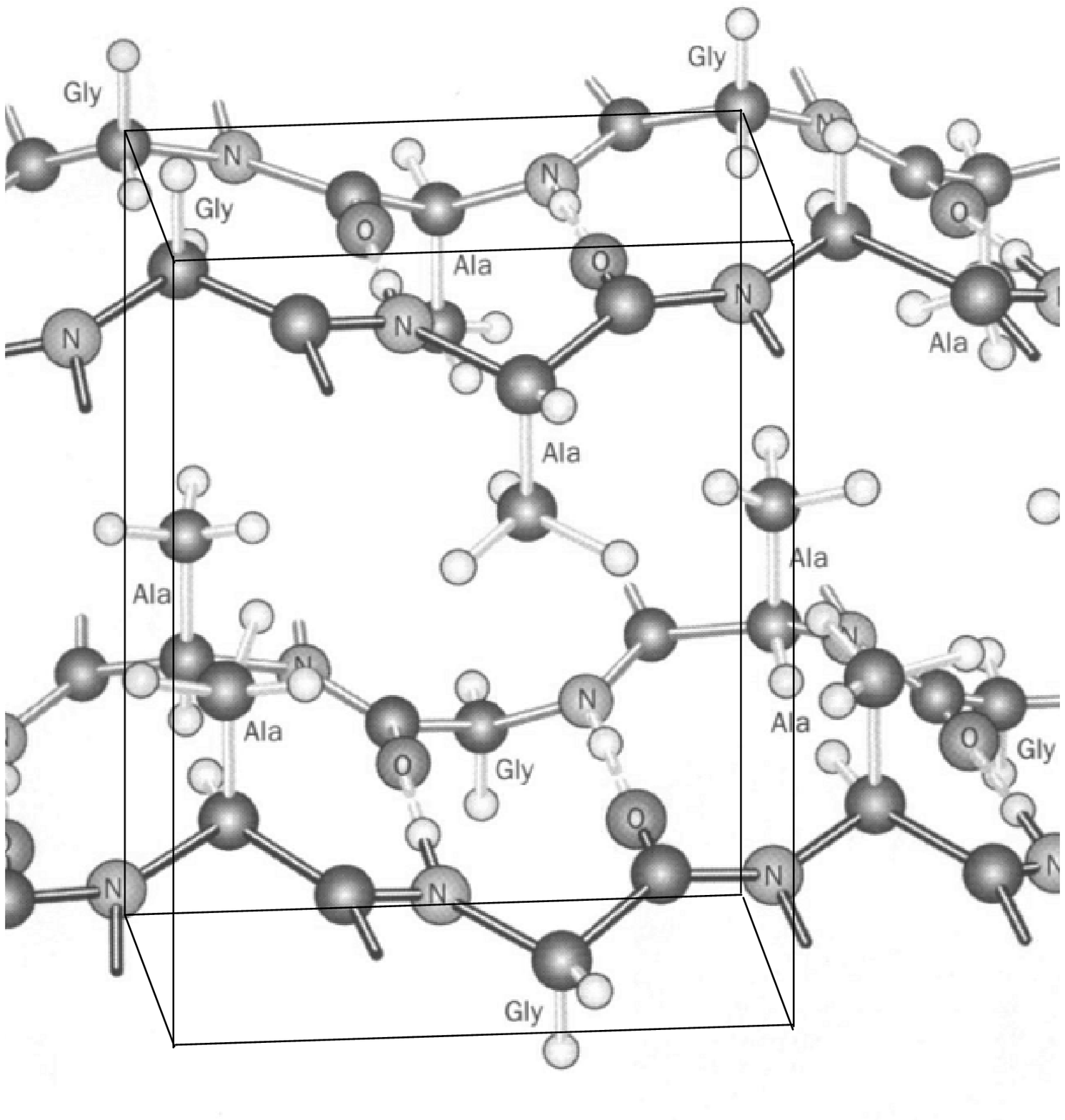}
   \centering
 \end{minipage}
 \caption{Top: Schematic view of the unit cell. Atom $k$ is
   loacted at position ${\bf r}_k$, and the atom type is specified by
   the form factor $f_k$ of the atom. For simplification schematic
   illustrations are in 2D, if possible, even though the model refers
   to three space dimensions, $D=3$. Bottom: Possible
   configuration inside the unit cell (illustration adapted from \cite{Zubay}).}
 \label{fig:UnitCell}
\end{figure}

In this section we present a simple model of spider silk which allows
us to compute the scattering function
\begin{equation}
 G({\bf q}) = \left\langle \left| \sum\nolimits_j 
f_j \exp(i {\bf q} {\bf r}_j) \right|^2 \right\rangle
\end{equation}
as measured in X-ray scattering experiments.  The atomic positions are
denoted by ${\bf r}_j$ and the atomic form factors by $f_j$. The
modeling proceeds on three different levels: 1) On the largest
lengthscales spider silk is modelled as an ensemble of crystallites
embedded in an amorphous matrix and preferentially
oriented along the fiber axis. 2) Each crystallite is composed of parallel or
antiparallel $\beta$-sheets. 3) Each unit cell contains a given number
of amino acids, whose arrangements have been classified by
Warwicker \cite{warwicker}. In Fig.~(\ref{fig:UnitCell}, bottom) we show an
illustration of Bombyx mori by Geis \cite{Zubay}. 

In the following we shall build up a model, starting on the smallest
scales and working up to the whole system.  Subsequently we will
compare our calculated scattering functions with experimental data.
Thereby we are able to determine the arrangement of atoms
in the unit cell which {\it optimizes} the agreement between model and
experiment.

\subsection{Unit cell}
One unit cell of a crystallite is described as a set of atoms at
\emph{positions} ${\bf r}_{k}$ relative to the center of the unit
cell, where $k = 1,2,...,K$ runs through the atoms of the unit cell
(see Fig.~\ref{fig:UnitCell}, top panel, for a schematic drawing). Each atom is
assigned a \emph{form factor} $f_k$, specifying the scattering
strength of the respective atom type.

\subsection{Crystallite}
\label{model:crys}
\begin{figure}[h]
 \includegraphics[width=6cm]{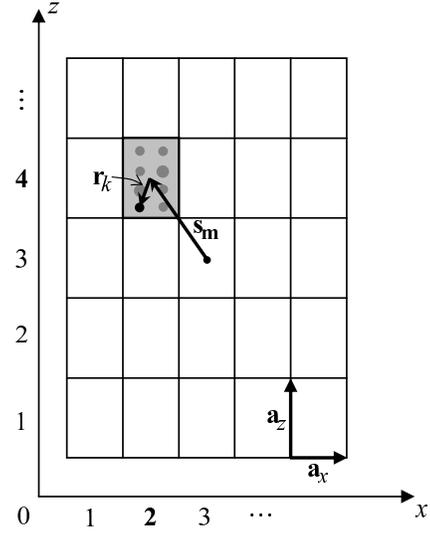}
 \centering
 \caption{Schematic view of the crystallite. The vector ${\bf s}_{\bf
     m}$ goes from the center of the crystallite to the unit cell
   ${\bf m}$. From there ${\bf r}_k$ goes to atom $k$.}
 \label{fig:crystallite}  
\end{figure}

A crystallite is composed of $M=M_x M_y M_z$ unit cells, replicated $M_x$, $M_y$
and $M_z$ times along the primitive vectors ${\bf a}_x$, ${\bf a}_y$
and ${\bf a}_z$, respectively. The unit cells in a crystallite
are numbered by a vector index ${\bf m} = (m_x, m_y, m_z)$, where
$m_\nu = 1,2,...,M_\nu$ for $\nu=x,y,z$. Hence the center of mass of
unit cell ${\bf m}$ has position vector
$$
{\bf\tilde{s}}_{\bf m} =
 m_x {\bf a}_x + m_y {\bf a}_y + m_z {\bf a}_z. 
$$
Actually it is more convenient to measure all distances with respect
to the center of the whole crystallite 
$${\bf s}_{\rm cm} = \frac{(M_x+1) {\bf a}_x + (M_y+1) {\bf
    a}_y + (M_z+1) {\bf a}_z}{2} $$
so that ${\bf{s}}_{\bf m}={\bf\tilde{s}}_{\bf m}-{\bf s}_{\rm cm}$
denotes the position of unit cell ${\bf m}$ relative to the center of
the crytallite to which it belongs (see Fig.~\ref{fig:crystallite})
and the position of atom $k$ in unit cell ${\bf m}$ relative to
the center of the crystallite is
\begin{equation}
 {\bf r}_{{\bf m},k} = {\bf s}_{\bf m} + {\bf r}_k 
\end{equation}

\subsection{Ensemble of Crystallites} \label{sec:Ensemble}
\begin{figure}[h]
  \includegraphics[width=7.5cm,angle=270]{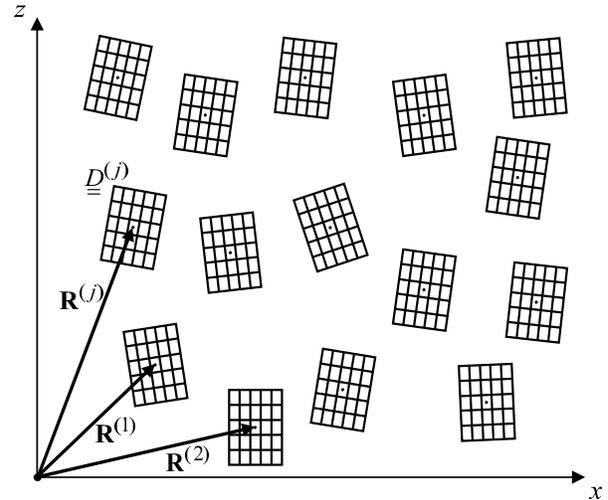}
  \centering
  \caption{The whole system is composed of crystallites at positions
    ${\bf R}^{(j)}$. They can be rotated by rotation matrices
    $\underline{\underline{D}}^{(j)}$.}
 \label{fig:system}  
\end{figure}

\begin{figure}[h]
 \begin{minipage}[t]{.45\textwidth}
   \includegraphics[width=6cm]{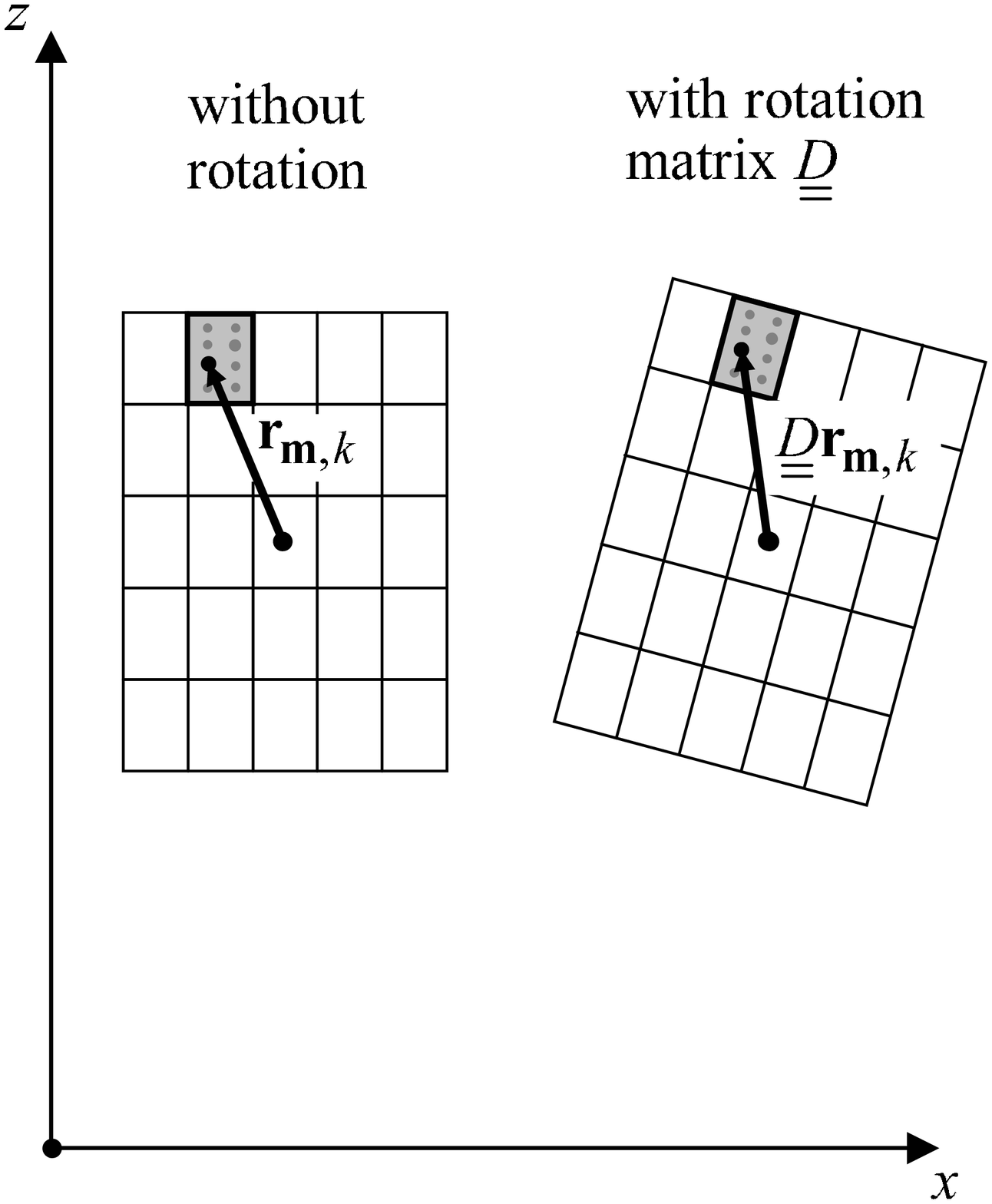}
   \centering
 \end{minipage}

 \begin{minipage}[t]{.45\textwidth}
   \includegraphics[width=4cm]{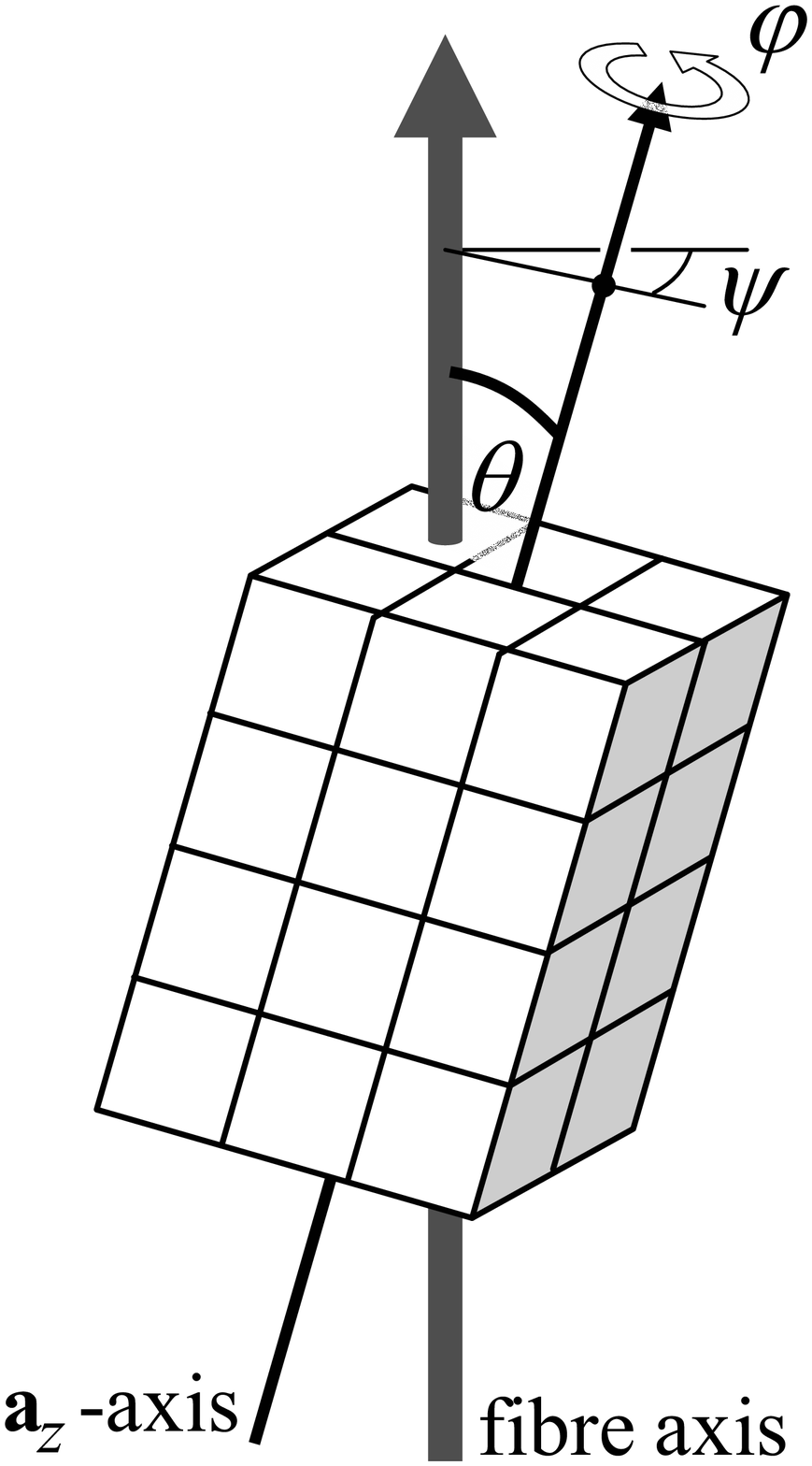}
   \centering
 \end{minipage}
 \caption{Top: The position of atom $k$ in unit cell ${\bf m}$ of
   the rotated crystallite is obtained by applying the rotation matrix
   $\underline{\underline{D}}{\bf r}_{{\bf m},k} $ to the position
   vector in the aligned configuration. Bottom:
   Illustration of Euler angles in 3D. The angle $\theta$ specifies
   the deviation of the crystallite's ${\bf a}_z$-axis from the fiber
   axis of the strand.  $\phi$ is the rotation of the crystallite
   about it's own ${\bf a}_z$-axis, and $\psi$ is the rotation of the
   ${\bf a}_z$-axis about the fiber axis (after the
   $\theta$-rotation).  }
 \label{fig:EulerAngles}  
\end{figure}

The whole system is composed of $N$ such crystallites at positions
${\bf R}^{(j)}$ with $j=1,2,...N$ (see Fig.~\ref{fig:system}).  The
crystallites are not perfectly aligned with the fiber axis, instead
their orientation fluctuates. The orientation of a single crystallite
is specified by three Euler angles $\phi^{(j)}, \theta^{(j)}, \psi^{(j)}$ (see
Fig.~\ref{fig:EulerAngles}, bottom). Here we have chosen the $z$-axis as
the fiber axis and $\theta$ denotes the angle between the
$z$-direction of the crystallites (direction of covalent bonds) and
the fiber axis. The atomic positions of the rotated crystallite are
obtained from the configuration which is perfectly aligned with the
$z$-axis by applying a rotation matrix $\underline{\underline{D}}^{(j)}$ (see
Fig.~\ref{fig:EulerAngles}, top):
\begin{equation}
 {\bf r}_{{\bf m},k}^{(j)} = {\bf R}^{(j)} +
 \underline{\underline{D}}^{(j)} 
{\bf r}_{{\bf m},k} \, .
\end{equation}

In the experiment the scattering intensity is obtained for a large
system, consisting of many crystallites. Hence it is reasonable to
assume that the scattering function is self-averaging and hence can be
averaged over the positions and orientations of the crystallites.
We use angular
brackets $\langle \,\, \rangle$ to denote the average of an observable
$\mathcal{O}$ over crystallite positions
${\bf R}^{(j)}$ and orientations $\underline{\underline{D}}^{(j)}$:
\begin{equation}
  \left\langle \mathcal{O} \right\rangle =
\int \prod_{j=1}^N \left(d^3{\! R}^{(j)} 
\mathcal{D}\underline{\underline{D}}^{(j)}
\right) \, 
\mathcal{P}_{\rm pos} ({\bf R}^{(1)},...,{\bf R}^{(N)}) 
\mathcal{O} \, .
\end{equation}
Here, the crystallite positions follow the distribution
function $\mathcal{P}_{\rm pos}({\bf R}^{(1)},...,{\bf R}^{(N)})$
which in general includes correlations. In contrast the orientation
of each crystallite is assumed to be independent of the others.
The average over all orientations
\begin{equation}
  \mathcal{D}\underline{\underline{D}}^{(j)} = 
d\phi^{(j)} d \theta^{(j)} d\psi^{(j)} \sin\theta^{(j)} \, 
 \mathcal{P}_{\rm angle}(\phi^{(j)}, \theta^{(j)}, \psi^{(j)}) 
\end{equation}
involves the angular distribution function 
$\mathcal{P}_{\rm angle}(\phi, \theta, \psi)$, which is the same for
each crystallite. 
In the simplest model we assume a Gaussian distribution for the
deviations of the crystallite axis from the fiber axis $\mathcal{P}_{\rm angle}(\phi, \theta, \psi)$ $\sim \exp(-\frac{\theta^2}{\theta_0^2})$
while all values of $\phi$ and $\psi$ between $0$ and $2\pi$ are
equally likely.

\subsection{Continuous background} 
\label{sec:background} 
\begin{figure}[h]
 \includegraphics[angle=270,width=8cm]{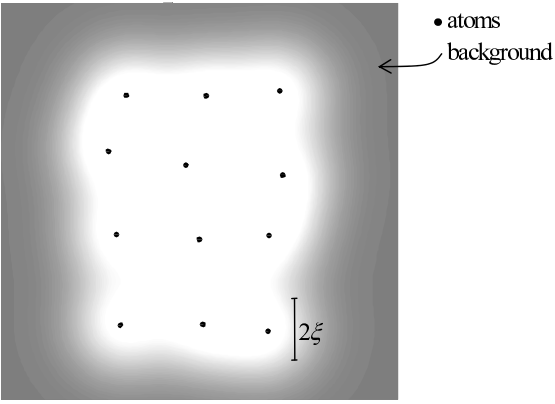}
 \centering
 \caption{Illustration of the continuous background, modelling the
   amorphous matrix. Outside the crystallites it has a homogeneous
   scattering density $\varrho_0$, which matches the mean scattering
   density of the crystallites. In this illustration $\xi$ is $0.45$ times the mean nearest neighbour distance.}
 \label{fig:background}  
\end{figure}

The space between the crystallites is filled with water molecules and
strands connecting the crystallites, which are called amorphous
matrix. In the scope of this work, we are not interested in the
details of its structure and model it as a continuous background
density $\varrho_0$, chosen to match the average scattering density of
the crystallite (see Fig.~\ref{fig:background}): $\varrho_0 = \sum_{k=1}^K f_k/V_{\rm uc}$. Here
$V_{\rm uc}$ is the volume of the unit cell. Inside the crystallite
there is no background intensity which is achieved in our model by
cutting out a spherical cavity, $V({\bf r})$, around each atom ${\bf
  r}={\bf r}_{{\bf m},k}^{(j)}$. For simplicity we assume a Gaussian
cavity
\begin{equation}
  V({\bf r}) = \frac{\bar f}{(2\pi\xi^2)^{3/2}} 
\exp\left( -\frac{{\bf r}^2}{2\xi^2} \right) \, .
\end{equation}
and choose the amplitude such that the average density inside the
crystallites is zero:
\begin{eqnarray}
  0 =     \int_{\rm crystallite} d^3  r \left(\varrho_0 - 
 \sum_{{\bf m}}^{\bf M} \sum_{k=1}^{K} V({\bf r} - 
{\bf r}_{{\bf m},k}) \right) \, ,
\end{eqnarray}
where the sum over the vector index $\sum_{{\bf m}}^{\bf M}$ means 
$\sum_{m_x=1}^{M_x}$ $\sum_{m_y=1}^{M_y}\sum_{m_z=1}^{M_z}$. 
With this assumption $\bar f=\sum_{k=1}^{K} f_k/K$ is simply the
\emph{average} form factor. The typical size of the cavity $\xi$ has
to be comparable to the nearest neighbour distance to make sure that
there is no ``background'' inside the crystallites. Models with and without continuous background are compared in Appendix \ref{sec:appBackground}.

This completes the specification of our model and we proceed to
compute the scattering function as predicted by the model.

\section{Scattering function\label{sec:sf}}
Given the atomic positions ${\bf r}_{{\bf m},k}^{(j)}$, the background
density $\varrho({\bf r})$ and the statistics of the crystallites'
orientations and positions, we calculate the scattering function:
\begin{eqnarray}
  && G({\bf q}) =   \nonumber  \\
&&\Biggl\langle \Biggl| \sum_{j=1}^N \sum_{{\bf m}}^{\bf M} 
\sum_{k=1}^{K} f_k \exp(i {\bf q}  {\bf r}_{{\bf m},k}^{(j)})  
+ \int
d^3 r \, \varrho({\bf r}) \exp(i{\bf q}  {\bf r})   \Biggr|^2 \Biggr\rangle \nonumber \\
\label{eq:G(q):def}
\end{eqnarray}
Here $\varrho({\bf r})$ is the background intensity whose Fourier
transform reads: 
\begin{eqnarray}
&& \int d^3 r \, \varrho({\bf r}) \exp(i{\bf q}{\bf r}) \nonumber \\
&=& \int d^3 r\exp(i{\bf q}{\bf r})\left(\varrho_0 - 
\bar{f} \sum_{j=1}^N \sum_{{\bf m}}^{\bf M} \sum_{k=1}^{K} 
V({\bf r}-{\bf r}_{{\bf m},k}^{(j)})\right)  \nonumber \\
&=& \varrho_0 V \delta_{{\bf q},{\bf 0}} - \tilde{V}({\bf q}) 
\sum_{j=1}^N \sum_{{\bf m}}^{\bf M} \sum_{k=1}^{K} 
\exp(i{\bf q}{\bf r}_{{\bf m},k}^{(j)}) \, 
\end{eqnarray}
The uniform density, giving rise to a contribution proportional to
$\delta_{{\bf q},{\bf 0}}$, does not contain information about the
structure of the system. Furthermore the central beam has to be
gated out in the analysis of the experimental data. Hence we neglect
the uniform contribution and obtain
for the scattering intensity (\ref{eq:G(q):def}) 
\begin{equation}
 G({\bf q}) = \left\langle \Bigg| \sum_{j=1}^N \sum_{{\bf m}}^{\bf M}
 \sum_{k=1}^{K} 
   \underbrace{\left(f_k - \tilde{V}({\bf q}) \right)}_{F_k ({\bf q})} 
 \exp(i {\bf q} \cdot {\bf r}_{{\bf m},k}^{(j)}) 
\Bigg|^2 \right\rangle \, .
\label{eq:G(q):def2} 
\end{equation}
The cavities give rise to effective form factors $F_k ({\bf q}) = f_k
- \tilde{V}({\bf q})$ accounting for the scattering of the atoms themselves,
$f_k$, and the cavities around them, $\tilde{V}({\bf q})$.
Note that, if we want to switch off the background $\bar{f} \to 0$, we
return to the original form factors $F_k ({\bf q}) \to f_k$.

Inserting the average $\langle \;\; \rangle$ and multiplying out the
magnitude squared in (\ref{eq:G(q):def2}) yields:
\begin{eqnarray}
\label{eq:G}
   G({\bf q}) = 
  &{\displaystyle\int}& \prod_{j=1}^N \left(d^3{\! R}^{(j)} \mathcal{D}\underline{\underline{D}}^{(j)}  \right) \,
  \mathcal{P}_{\rm pos}({\bf R}^{(1)},...,{\bf R}^{(N)}) \nonumber \\ 
&\times& \sum_{j,j'=1}^N \exp\left(i {\bf q} ( {\bf R}^{(j)} - {\bf R}^{(j')} ) \right) \nonumber \\
&\times& \sum_{{\bf m},{\bf m}'}^{\bf M} \sum_{k,k'=1}^{K} F_k ({\bf q}) F_{k'}^* ({\bf q}) \nonumber \\
&\times&  \exp \!\left(i{\bf q} ( \underline{\underline{D}}^{(j)} 
   {\bf r}_{{\bf m},k} - \underline{\underline{D}}^{(j')} {\bf
     r}_{{\bf m}',k'}) 
\right)
\end{eqnarray}
Note, that the double sum over the crystallites $j$ and $j'$ also
applies to the rotation matrices $\underline{\underline{D}}^{(j)}$ and
$\underline{\underline{D}}^{(j')}$ in the second line. We now split
the scattering function in two terms $G({\bf q}) = G_1({\bf q}) +
G_2({\bf q})$: The first one, $G_{1}({\bf q})$, only includes the
terms $j=j'$ of that sum and thus incorporates scattering of the same
crystallite, but not scattering from different crystallites. The second
one, $G_{2}({\bf q})$, only including the terms with $j \ne j'$, takes
into account coherent scattering of two different crystallites.

\subsection{Incoherent part}
We first consider the case $j=j'$ of the sum, i.e. the contribution to
the scattering function which is incoherent with respect to different
crystallites. Here, the term $\exp(i {\bf q} ( {\bf R}^{(j)} - {\bf
  R}^{(j')} ) )$ gives $1$, and therefore the integral over the
crystallite positions ${\bf R}^{(j)}$ can be performed and trivially
gives 1 due to the normalization of the spatial distribution
$\mathcal{P}_{\rm pos}({\bf R}^{(1)},...,{\bf R}^{(N)})$.
Furthermore, for each summand $j$, all integrations over the
orientations $\underline{\underline{D}}^{(1)}, ...,
\underline{\underline{D}}^{(N)}$, except
$\underline{\underline{D}}^{(j)}$, can be performed and also yield 1.
Therefore, the $N$ terms with $j=j'$ simplify to:
\begin{eqnarray}
 G_{1}({\bf q}) &=& N \int \mathcal{D}\underline{\underline{D}} \, 
\sum_{{\bf m},{\bf m}'}^{\bf M} \sum_{k,k'=1}^{K} F_k ({\bf q}) F_{k'}^* ({\bf q}) \nonumber \\
 &&\qquad \qquad \qquad \times \exp \!\left(i {\bf q} ( \underline{\underline{D}} {\bf r}_{{\bf m},k} 
- \underline{\underline{D}} {\bf r}_{{\bf m}',k'}) \right) \nonumber \\
 &=& N \int \mathcal{D}\underline{\underline{D}} \left| 
\sum_{{\bf m}}^{\bf M} \sum_{p=1}^{P} F_k ({\bf q}) 
\exp \!\left(i (\underline{\underline{D}}^T {\bf q})  
\cdot {\bf r}_{{\bf m},k} \right) \right|^2 \nonumber \\
 &=& N \int \mathcal{D}\underline{\underline{D}} \left| 
A(\underline{\underline{D}}^T {\bf q})  \right|^2,
\end{eqnarray}
where $\underline{\underline{D}}^T$ is the transpose of the matrix
$\underline{\underline{D}}$ and
\begin{equation}
 A({\bf q}) := \sum_{{\bf m}}^{\bf M} \sum_{k=1}^{K} 
F_k ({\bf q}) \exp(i {\bf q} \cdot {\bf r}_{{\bf m},k}) \label{eq:A(q)}
\end{equation}
is the unaveraged scattering amplitude of a single unrotated
crystallite. Hence the interpretation of the incoherent part
$G_{1}({\bf q})$ is straightforward: Each crystallite contributes
independently, each with a given orientation. The orientation can be
absorbed in the scattering vector ${\bf q}\to
\underline{\underline{D}}^T {\bf q}$, so that the contributions of two
crystallites with different orientations are simply related by a
rotation of the scattering vector.
In the macroscopic limit we are allowed to average
over all possible orientations and get a sum of $N$ identical terms.

\subsection{Coherent part}
We now consider the contribution to the scattering intensity from \emph{different} crystallites, i.e.~the case $j \ne j'$ in Eq.(\ref{eq:G}). In
analogy to the above calculation all integrations over the orientations
$\underline{\underline{D}}^{(1)}, ...,
\underline{\underline{D}}^{(N)}$, can be performed except for
$\underline{\underline{D}}^{(j)}$ and
$\underline{\underline{D}}^{(j')}$:
\begin{eqnarray}
  G_{2}({\bf q}) &=& 
 \int d^3 \!R^{(1)} \cdots d^3 \!R^{(N)} \, \mathcal{P}_{\rm pos}({\bf R}^{(1)},...,{\bf R}^{(N)}) \nonumber \\ 
&\times& \sum_{j \ne j'} \exp \! \left(i {\bf q} ( {\bf R}^{(j)} - {\bf R}^{(j')} ) \right) \nonumber \\
&\times& \int \mathcal{D}\underline{\underline{D}} \, \mathcal{D}\underline{\underline{D}}' \sum_{{\bf m},{\bf m}'}^{\bf M} 
\sum_{k,k'=1}^{K} F_k ({\bf q}) F_{k'}^* ({\bf q}) \nonumber \\
&\times& \exp \!\left(i{\bf q} ( \underline{\underline{D}} {\bf r}_{{\bf m},k} 
- \underline{\underline{D}}' {\bf r}_{{\bf m}',k'}) \right) 
 \label{eq:G2}
\end{eqnarray}
where we have used that the angular distribution is the same for all
crystallites.
We introduce the structure factor of the crystallite positions:
\begin{eqnarray}
 S({\bf q}) &=& \frac 1 N \left\langle \sum_{j,j'=1}^N \exp \!\left( i {\bf q} ( {\bf R}^{(j)} - {\bf R}^{(j')} ) \right) \right\rangle \nonumber \\
	    &=& 1 + \frac 1 N \int d^3 \!R^{(1)} \cdots d^3 \!R^{(N)} \, \mathcal{P}_{\rm pos}({\bf R}^{(1)},...,{\bf R}^{(N)}) \nonumber \\
            &&\qquad \qquad   \times \sum_{j \ne j'} \exp\!\left(i {\bf q} ( {\bf R}^{(j)} - {\bf R}^{(j')} ) \right) \,
\end{eqnarray}
and observe that the two upper lines in (\ref{eq:G2}) are just 
$N \left( S({\bf q}) - 1 \right)$. Hence $G_2$ can be simplified to:
\begin{eqnarray} 
G_{2}({\bf q}) &=& 
 N \left(S({\bf q}) - 1 \right) \nonumber \\
  && \times \left|  \int \mathcal{D}\underline{\underline{D}} \, \sum_{{\bf m}}^{\bf M} \sum_{k=1}^{K} F_k ({\bf q}) \exp \!\left(i (\underline{\underline{D}}^T {\bf q}) \cdot {\bf r}_{{\bf m},k}  \right) \right|^2 \nonumber \\
 &=&  N \left(S({\bf q}) - 1 \right) 
\left|  \int \mathcal{D}\underline{\underline{D}} \, A(\underline{\underline{D}}^T {\bf q})  \right|^2
\end{eqnarray}
Again the interpretation is straightforward: For coherent scattering
the amplitudes of individual crystallites with different orientations
add up, as expressed by $ \int \mathcal{D}\underline{\underline{D}} \,
A(\underline{\underline{D}}^T {\bf q}) $. Spatial correlations of the
centers of the crystallites are accounted for by the structure
function.

The total scattering function 
\begin{equation} \label{eq:ScatteringFunctionFinal}\fbox{\parbox{0.49\textwidth}{
\[ \frac{G({\bf q})}{N} = 
\int \mathcal{D}\underline{\underline{D}} \left| A(\underline{\underline{D}}^T {\bf q})  \right|^2
+
\left(S({\bf q}) - 1 \right) \left|  \int \mathcal{D}\underline{\underline{D}} \, A(\underline{\underline{D}}^T {\bf q})  \right|^2\]}}
\end{equation}
is reduced to the scattering amplitude of a
{\it single crytallite} $A({\bf q})$, which we compute next. Note that if
the angular spread of the crystallites can be neglected, in other
words all crystallites are approximately aligned, then the above
expression reduces to $G({\bf q})=N S({\bf q}) \left| A({\bf q})
\right|^2$, as one would expect.

If there is thermal motion of the atoms around their equilibrium
positions due to finite temperature, the intensity of the scattering function
$G({\bf q})$ is weakened for larger $q$-values
\cite[e.g.][]{willis75}. The resulting scattering function has to be
multiplied with the Debye-Waller factor
\begin{equation}
 G_{\text DW}({\bf q}) = G({\bf q}) \cdot \exp(- {\bf q}^2 \langle u^2 \rangle) \,, \label{eq:Debye-Waller}
\end{equation}
where $\langle u^2 \rangle$ is the mean square displacement of the atoms in any direction.

\subsection{Scattering amplitude of a single crystallite} 
\label{sec:A(q)}
The calculation of the scattering amplitude of a single crystallite
$A({\bf q})$ (defined in Eq.~\ref{eq:A(q)}) follows standard procedures. We
substitute the atomic positions of Sec.~\ref{model:crys} 
and note that the sums over $m_x$, $m_y$ and $m_z$ are geometric
progressions which can easily be performed, yielding: 
\begin{equation} \label{eq:ScatteringAmplitude}
A({\bf q}) = L_{M_x}({\bf q} {\bf a}_x) L_{M_y}({\bf q} {\bf a}_y) 
L_{M_z}({\bf q} {\bf a}_z) \sum_{k=1}^{K} F_k({\bf q}) \exp(i {\bf q} {\bf r}_{k} ) \, . 
\end{equation}
Here $L_{M_\nu} ({\bf q} {\bf a}_\nu) = \frac{\sin({\bf q} {\bf a}_\nu
  M_\nu / 2)} {\sin( {\bf q} {\bf a}_\nu / 2)}$ is the well-known Laue
function, which has an extreme value, when its argument ${\bf q} {\bf
  a}_\nu$ is a multiple of $2\pi$. It is noteworthy, however, that the
position of the extremum of the magnitude of the \emph{scattering
  amplitude} $A({\bf q})$ may be shifted, if the form factor of the
unit cell $\sum_{k=1}^{K} F_k({\bf q}) \exp(i {\bf q} {\bf r}_{k} )$ has a
non-vanishing gradient at that position and $M_\nu$ is finite (and
therefore the peak width of the Laue function is nonzero). In this
case the resultant peak may be shifted by a value of the order of its
peak width. Consequently care has to be taken, when determining 
lattice constants from experimental peak positions.

\section{Atomic configuration of the unit cell}
\label{sec:ConfigurationUnitCell} 
The computation of the scattering
function $G({\bf q})$ requires the atomic configuration $\{ {\bf r}_k
\}_{k=1}^K$ of the unit cell, which we discuss next.

\subsection{Unshifted unit cells} \label{sec:unshifted} 
\begin{figure}[h]
 \includegraphics[width=5cm]{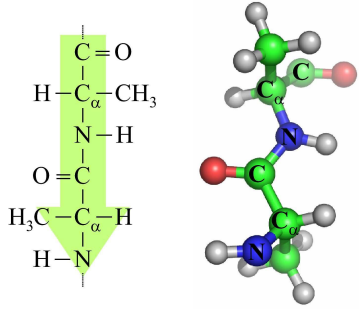}  
 \centering
 \caption{(Color online) Chemical structure (left) and conformation
   (right) of a poly-alanine strand. Two alanines are shown. The
   CH$_3$ group is characteristic for the alanine amino acid and is
   bound to the so called C$_\alpha$ atom. The arrow indicates the
   direction ${\rm C}\rightarrow{\rm C}_\alpha\rightarrow{\rm N}$ of
   the backbone.}
 \label{fig:AlaninChem}  
\end{figure}

It is known
that the crystallites are composed of poly-alanine strands (see
\cite{arnott67,grubb97}). In Fig.~\ref{fig:AlaninChem}, two alanine
amino acids of this strand are shown. Its conformation, shown on the
right side, is well established and was created with Yasara \cite{meling2006}. There are
two constraints for the strand: First, the subsequent alanines in the
strand must have the same orientation so that the strand does not have
a ``twist'' and can produce periodic structures. Second, the distance
between adjacent alanines has to match the size of the unit cell.
These two constraints allow for a unique choice of the two degrees of
freedom, namely the Ramachandran angles \cite{ramachandran} $\Phi$ and $\Psi$ which
are the dihedral angles for the bonds C$_\alpha$-N and C$_\alpha$-C,
respectively.

\begin{figure*}[h]
\centering
 \begin{minipage}[h]{6cm}
   \includegraphics[bb=0 0 1066 1464,height=5cm]{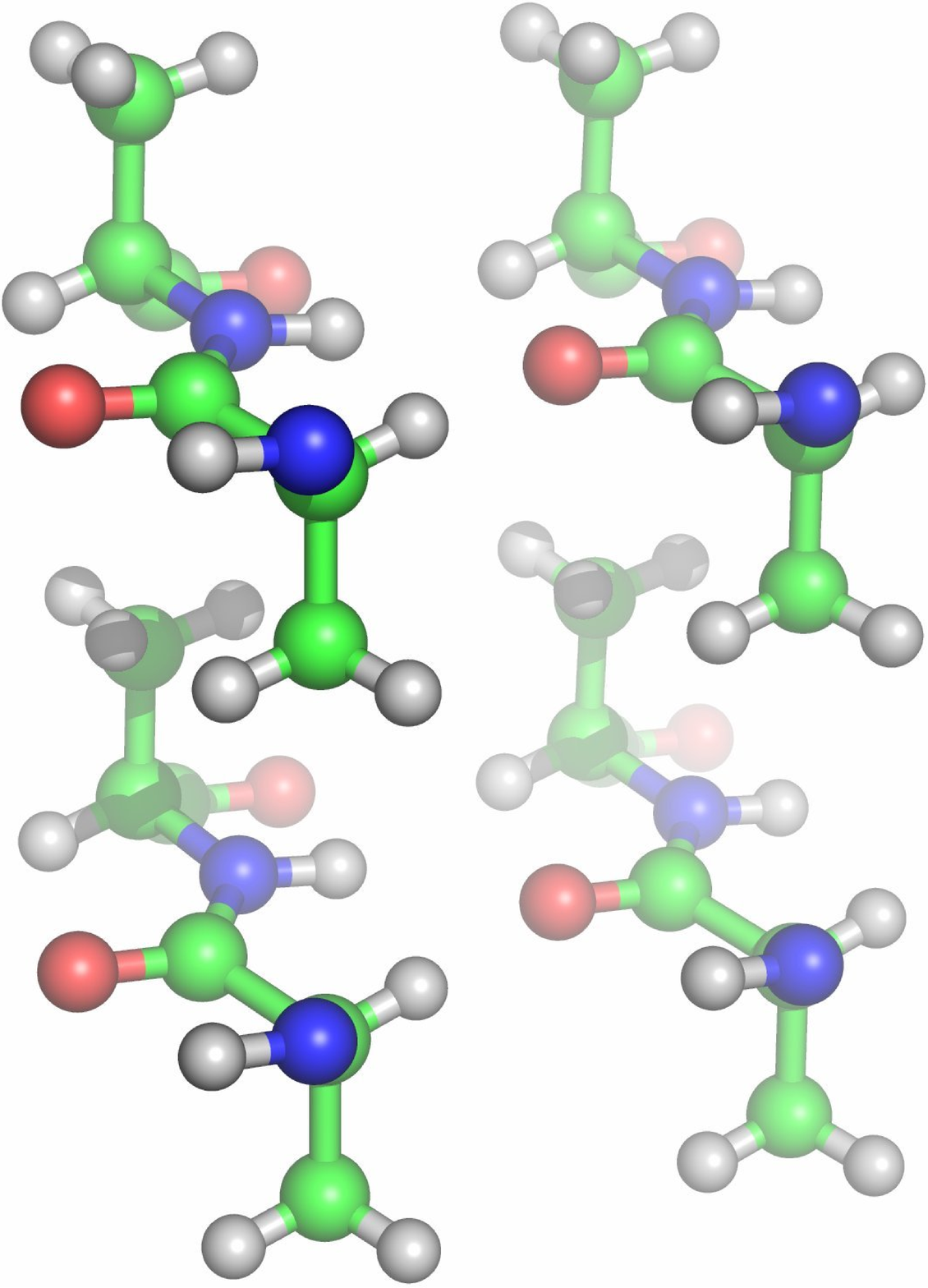}
   \centering
 \end{minipage}
 \begin{minipage}[t]{1cm}
   \includegraphics[width=1cm]{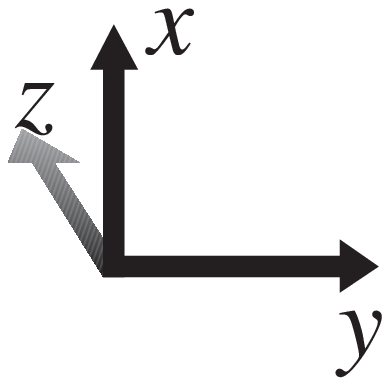}
   \centering
 \end{minipage}
 \begin{minipage}[h]{6cm}
   \includegraphics[bb=0 0 1284 1464,height=5cm]{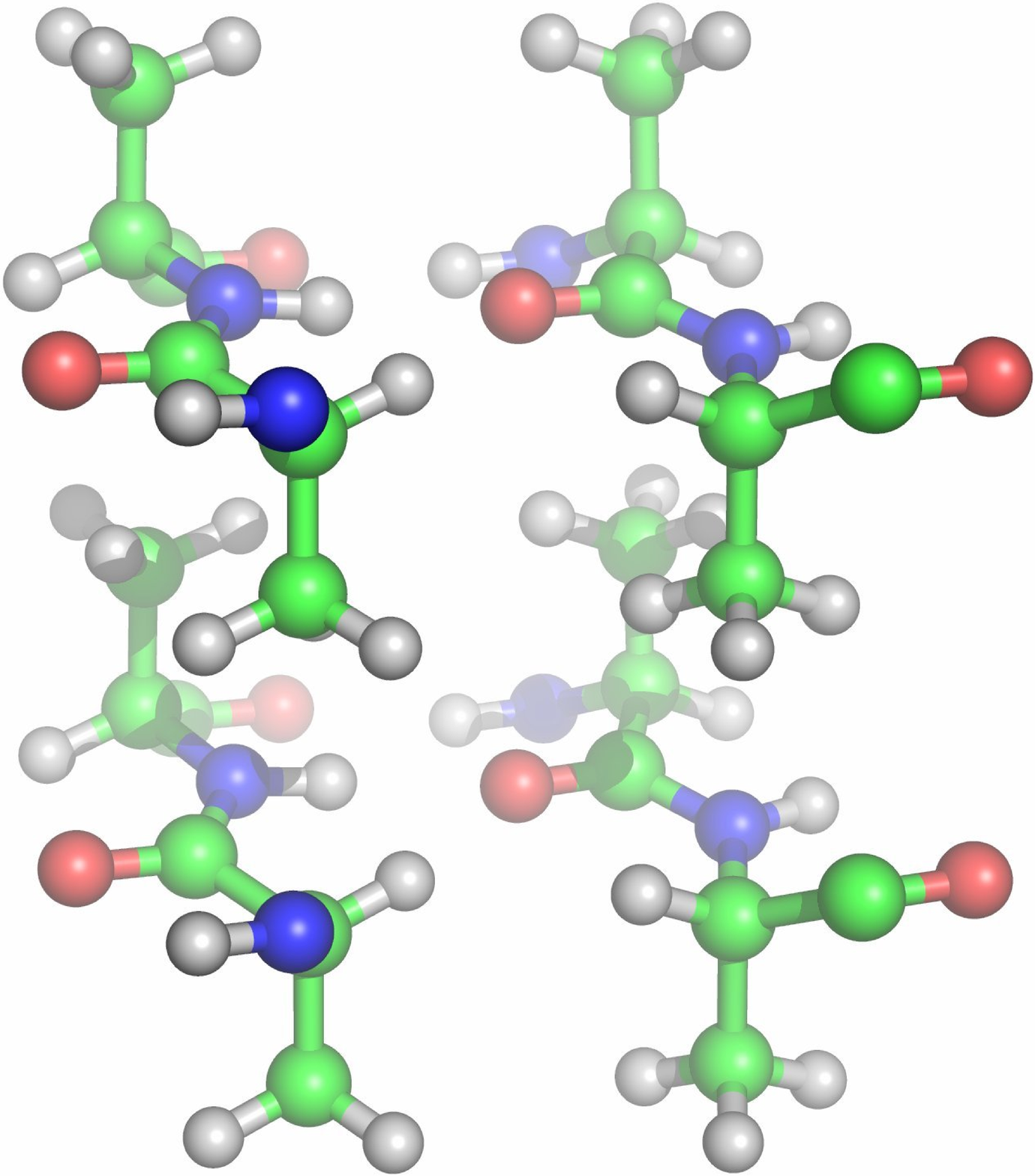}
   \centering
 \end{minipage}

\begin{minipage}[h]{.95\textwidth}
  \includegraphics[width=.77\textwidth]{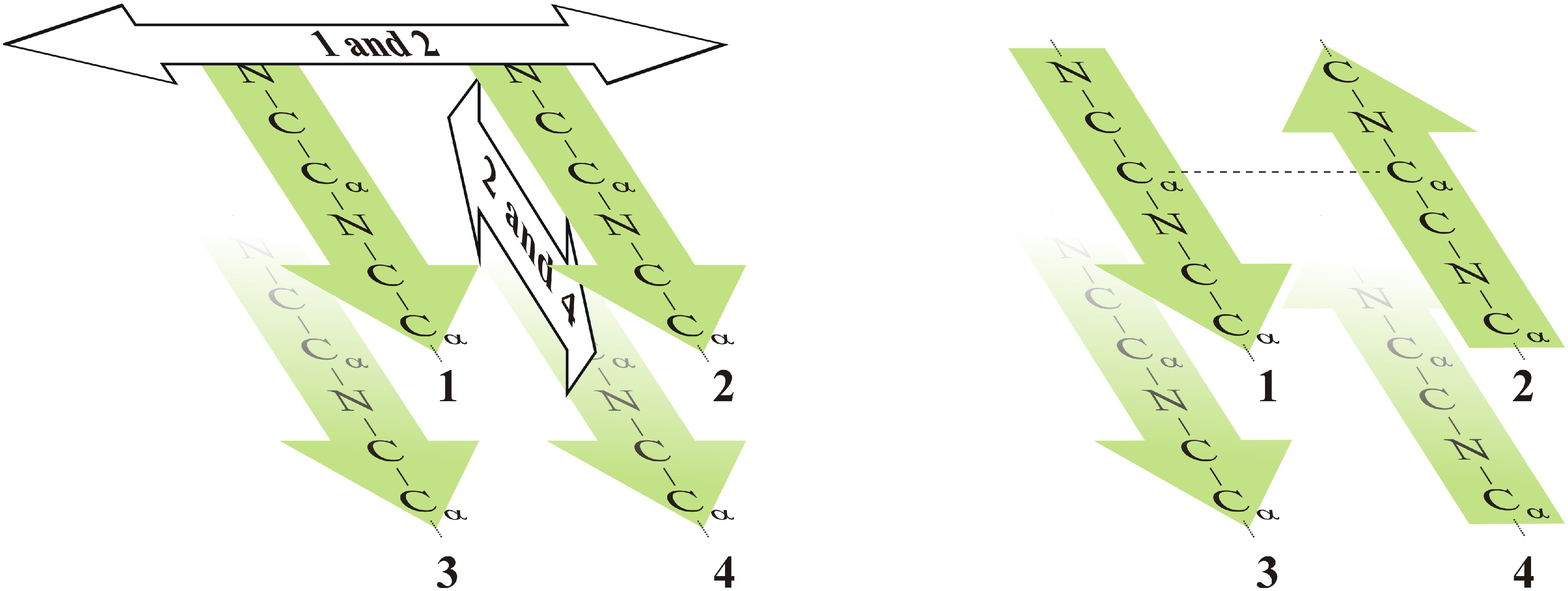}
  \centering
\end{minipage}

\vspace{.75cm}

\begin{minipage}[h]{\textwidth}
  \includegraphics[width=.7\textwidth]{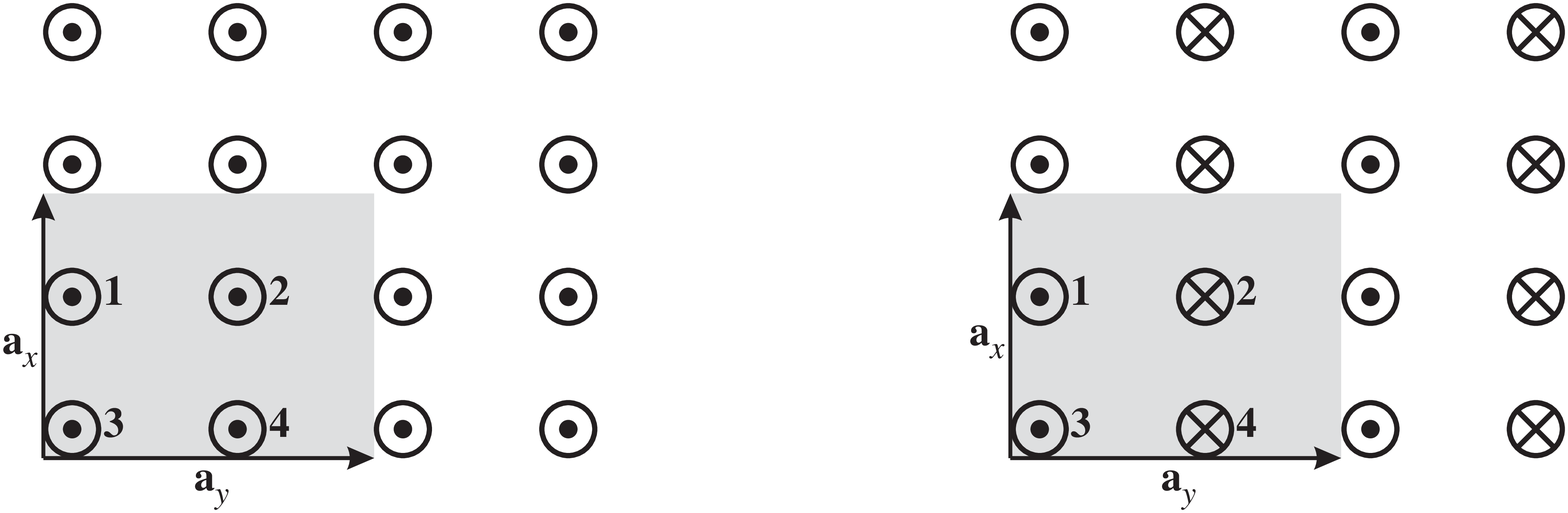}
  \centering
\end{minipage}

   \caption{(Color online) Illustration of one unit cell, containing
     four neighboring poly-alanine strands. Top: configuration of the
     atoms. Center: Schematic view displaying only the backbone atoms.
     The arrows illustrate two possible adjustments of the structure,
     which are to be optimzied to match the experimental scattering
     image. Bottom: Simplified schematic view along the backbone axis
     ${\bf a}_z$, where $\odot$ indicates an arrow pointing towards
     the reader and $\otimes$ pointing away from the reader. Here the
     unit cell is shown as a gray rectangle. In each case, a parallel
     structure is shown on the left side, and an antiparallel
     structure on the right side. The illustration shows an
     \emph{unshifted} configuration, that means: The C$_\alpha$ atoms
     of neighboring strands are aligned and have no shift in
     $z$-direction (indicated as dashed line in the middle right
     image). Furthermore, the strands are exactly aligned in the $x$-
     and $y$-direction.}
 \label{fig:AlaninUnitCell}
\end{figure*}

Many of the described poly-alanine strands side by side form a stable
crystalline configuration, the \emph{$\beta$-pleated-sheet}.  We
assume an orthorombic unit cell\footnote{We note that the assumption
  of an orthorhombic unit cell is restrictive. In a more general
  approach one can give up this assumption and use a smaller unit cell
  allowing for different shifts. The best fit to the data is obtained
  for a shift which can also be achieved with an orthorhombic unit
  cell. For the sake of clarity, we stick to the established
  unit cell notations and indexing of the reflexes here.}
, consisting of four alanine strands,
as illustrated in Fig.~\ref{fig:AlaninUnitCell}. Thus one unit cell
contains 8 alanine amino acids. Furthermore we define the spatial
directions in the usual way \cite[e.g.][]{warwicker}:
\begin{itemize}
\item[$x$:] Direction of the CH$_3$-groups and Van-der-Waals
  interactions between sheets lying upon each other.
\item[$y$:] Direction of the hydrogen bonds between the O-atom of one
  strand and the H-atom of the neighboring strand.
 \item[$z$:] Direction of the covalent bonds along the backbone.
\end{itemize}
Accordingly, ${\bf a}_x$, ${\bf a}_y$ and ${\bf a}_z$ are the
principal vectors pointing in these directions and $a_x$, $a_y$, $a_z$
their magnitudes. Note furthermore that due to symmetry the distance
between the strands has to be $a_x/2$ in the $x$-direction and $a_y/2$
in the $y$-direction.

In general one has to distinguish between the \emph{parallel} and
\emph{antiparallel} structure. In the parallel structure the
direction of the atom sequence ${\rm C}\rightarrow{\rm
  C}_\alpha\rightarrow{\rm N}$ in the strand's backbone is the same
for all strands (left side of Fig.~\ref{fig:AlaninUnitCell}). For the
antiparallel structure, this direction is alternating along the
$y$-axis (right side of fig.~\ref{fig:AlaninUnitCell}). Both will be
considered in the following analysis.

\subsection{Possible shifts inside the unit cell}
\label{shifts}
\begin{figure*}[h]
   \includegraphics[width=.8\textwidth]{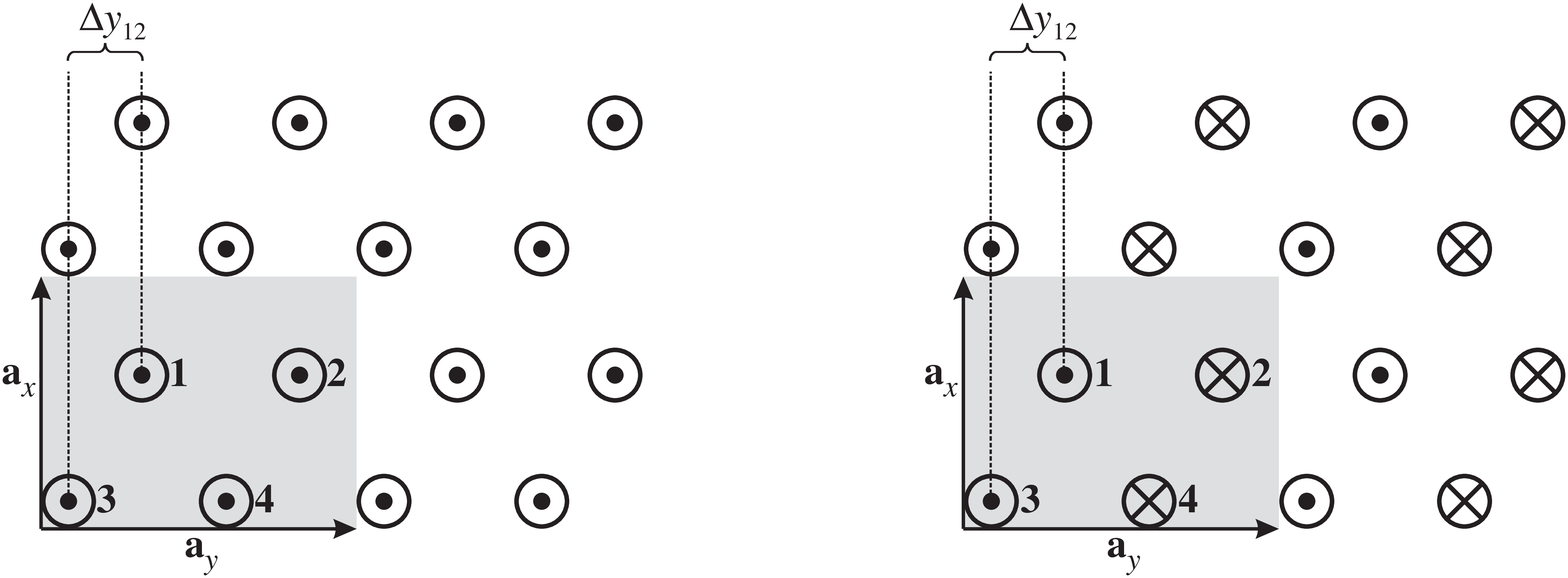}
   \centering
   \caption{Illustration of the unit cell, as at the bottom of
     Fig.~\ref{fig:AlaninUnitCell}, but with the conducted shift
     $\Delta y_{12} = a_y/4$ of strands 1 and 2 the in $y$-direction.}
 \label{fig:AlaninShifts}
\end{figure*}

The scattering intensity is not only sensitive to the conformation of
the poly-alanine strands, but also to the distance and orientation of
different strands relative to each other. Besides the question of
parallel or antiparallel structure, the four strands can also be
shifted with respect to each other. In principle there are four ways
to displace them (see Fig.~\ref{fig:AlaninUnitCell}):
\begin{itemize}
\item Shifting strands 1 and 2 in the $y$-direction by a value $\Delta
  y_{12}$. This displacement is performed in Fig.~\ref{fig:AlaninShifts}.
\item Shifting strands 1 and 2 in the $z$-direction by a value $\Delta
  z_{12}$. Because of the CH$_3$-groups extending into the layers
  above and below (as seen in Fig.~\ref{fig:AlaninUnitCell}, top panel), a
  displacement like this is only possible, if those two strands have a
  shift in the $y$-direction of $\Delta y_{12} \approx \pm {\bf
    a}_y/4$, as well. In this case the CH$_3$-groups can pass each
  other without overlapping.
 \item Shifting strands 2 and 4 in the $z$-direction by a value $\Delta z_{24}$. This shift was originally suggested by Arnott \emph{et al.} \cite{arnott67}.
 \item Shifting strands 2 and 4 in the $x$-direction. However this
   displacement would have a high energy cost, because it would break
   the hydrogen bonds between the H- and O-atoms of neighboring
   strands. Since, moreover, no reasonable result could be achieved
   performing such shifts, it will not be included in our discussion
   any further.
\end{itemize}

Note that shifting strands 1 and 2 in the $x$-direction or strands 2
and 4 in the $y$-direction is associated with resizing the unit cell
in the $x$- or $y$-direction, respectively.

\subsection{Variations between crystallites\label{sec:crystVariations}}
For real systems, the composition of the crystallite is certainly not fixed, but may vary from crystallite to cyrstallite. These variations can easily be included into our model introducing a probability distribution $P(\{r_{{\bf m},k}\})$ for the crystallite configuration $\{r_{{\bf m},k}\}$, with the normalization $\sum_{\{r_{{\bf m},k}\}} P(\{r_{{\bf m},k}\}) = 1$. The calculation is performed completely analogous to orientational distribution. The scattering function becomes
\begin{eqnarray}
 \label{eq:ScatteringFunctionVar}
 \frac{G({\bf q})}{N} &=& 
\sum_{\{r_{{\bf m},k}\}} P(\{r_{{\bf m},k}\}) \int \mathcal{D}\underline{\underline{D}} \left| A(\underline{\underline{D}}^T {\bf q})  \right|^2 \\
&+&
\left(S({\bf q}) - 1 \right) \left| \sum_{\{r_{{\bf m},k}\}} P(\{r_{{\bf m},k}\}) \int \mathcal{D}\underline{\underline{D}} \, A(\underline{\underline{D}}^T {\bf q})  \right|^2 \,. \nonumber
\end{eqnarray}

Later (sec.~\ref{sec:resultsModel}) we will investigate the effect of variable crystallite sizes as well as the influence of small fractions of glycine inside the crystallites.

\section{Results\label{sec:results}}
\subsection{Experimental scattering function}

Two types of samples have been investigated: fiber bundles and single fiber
preparations. Single fibers demand a highly collimated and brilliant beam, but are
better defined in orientation and are amenable tosimultaneous strain-stress measurements. 

An oriented bundle of major ampullate silk (MAS) of Nephilia Clavipes
was measured at the D4 bending magnet of HASYLAB/DESY in Hamburg. The
fibers were reeled on a steel holder and oriented horizontally in the
beam. The estimated number of threads was 400-600. Photon energy was
set to $E=10.9$ keV by a Ge(111) crystal monochromator located behind
a mirror to suppress higher harmonics. Data was collected with a CCD
x-ray camera (SMART Apax, AXS Bruker) with 60$\,\mu$m pixel size and
an active area of $62\,{\rm cm} \times 62\,{\rm cm}$. Illumination was
triggered by a fast shutter. The momentum transfer was calibrated by a
standard (corrundum). Raw data was corrected by empty image
(background subtracted). 

The single fiber experiments have been carried out at the microfocus beamline ID13
at ESRF, Grenoble \cite{id13}. A 12.7 keV x-ray beam was focused
with a pair of short focal length Kirkpatrick Baez (KB)
mirrors \cite{KB} to a $7\mu m$ spot at the sample. This focusing
scheme provides a sufficient flux density ($6.8 \cdot 10^{15}cps/mm^2$) to
obtain diffraction patterns from single dragline fibers.
The single fiber diffraction patterns were recorded with a CCD
detector positioned 131 mm behind the sample  (Mar 165
detector, Mar USA, Evanston, IL). One
of the beamline's custom made lead beamstops (approx. 300$\mu m$
diameter) was used to block the intense primary beam.
The raw data was treated as follows: (i) both the image and the
background (empty beam) were corrected by dark dark current,
and (ii) the (empty beam) background was subtracted from the image.
The peaks which are significantly broadened by the small
crystallite size (see below) can then be indexed to the orthorombic
lattice described above. Typical scattering distribution for both types of sample preparations
are shown in fig.~\ref{fig:experimentsalditt} as a function of parallel and
vertical momentum transfer.  More details on experimental procedures
and on the sample preparation by forced silking can be found in
\cite{GlisovicAPA07,Glisovic08}.

\subsection{Scattering function from the model\label{sec:resultsModel}}

It is our aim to determine those crystallites' parameters, which best
match the experimental result. The free parameters of our model are
the three shifts $\Delta y_{12}$, $\Delta z_{12}$ and $\Delta z_{24}$,
the unit cell dimensions $a_x$, $a_y$ and $a_z$, the crystallite size
in the three directions $M_x$, $M_y$ and $M_z$, as well as $\theta_0$,
the tilting angle of the crystallites away from the fiber axis.

The parameters of our model affect the scattering intensity in
different ways, which allows us to at least partially separate the
effects of different parameters. The crystallite size $(M_x,M_y,M_z)$
determines the peak widths, whereas the length of the principal
vectors ${\bf a}_x$, ${\bf a}_y$ and ${\bf a}_z$ determine the peak
position. (We have to keep in mind, however, that the peak position
can differ from the extremal values of the Laue functions, as
explained in section \ref{sec:A(q)}.)  The shifts $\Delta y_{12}$,
$\Delta z_{12}$ and $\Delta z_{24}$, as described in Sec.~(\ref{shifts}), 
affect the relative peak intensities via the form
factors of the unit cell, $\sum_{k=1}^{K} F_k({\bf q}) \exp(i {\bf q}
{\bf r}_{k})$.  Finally, the parameter $\theta_0$ is responsible for
the peak widths in the \emph{azimuthal} direction on the scattering
image.

\begin{table*}[t]
 \centering
 \begin{tabular}{|l|l|l|l|l|l|}
 \hline
             & \multicolumn{2}{l|}{presented calculation} & Warwicker \cite{warwicker54} & Marsh \cite{marshShort}& Arnott \cite{arnott67}   \\
 
 structure of & \multicolumn{2}{l|}{Nephila clavipes}    & Bombyx mori                   &  Tussah Silk    & poly-L-alanine   \\
 \hline
 alignement & parallel     & anti-parallel            & anti-parallel                 &  anti-parallel   & anti-parallel\\
 \hline\hline
 $a_x$ & $10.0 \; \text{\AA}$ & $10.0 \; \text{\AA}$ & $10.6 \; \text{\AA}$&$10.6 \; \text{\AA}$& $10.535 \; \text{\AA}$\\
 $a_y$ & $9.3 \; \text{\AA}$  & $9.3 \; \text{\AA}$ & $9.44 \; \text{\AA}$&$9.44 \; \text{\AA}$& $9.468 \; \text{\AA}$\\
 $a_z$ & $6.95 \; \text{\AA}$ & $6.95 \; \text{\AA}$ & $6.95 \; \text{\AA}$&$6.95 \; \text{\AA}$& $6.89 \; \text{\AA}$\\
 \hline
 $M_x$ & 1.5 $^{(*)}$         & 1.5 $^{(*)}$         &-                    &-                    &-\\
 $M_y$ & 5		      & 5                    &-                    &-                    &-\\
 $M_z$ & 9		      & 9                    &-                    &-                    &-\\
 \hline
 $\Delta y_{12}$ & $a_y/4$   & $a_y/4$               &0                    &$a_y/4$             &$\pm a_y/4$ $^{(**)}$\\
 $\Delta z_{12}$ & $0$	     & $0$                   &0                    &0                   &0\\
 $\Delta z_{24}$ & $0$	     & $-a_z/6$              &0                    &0                   &$-a_z/10$ \\
 \hline
 $\theta_0$    & $7.5^\circ$ & $7.5^\circ$           &-                    &-                   &-\\
 $\langle u^2 \rangle$ & 0.1\,\AA$^2$ &0.1\,\AA$^2$  &-                    &-                   &-\\
 \hline
 \end{tabular}
 \caption{Summary of parameters. The left two columns show the best match between
   experimental and calculated scattering functions. For $a_z = 6.95 \; \text{\AA}$, the resulting Ramachandran angles are
   $\Phi = -139.0^\circ$ and $\Psi = 136.9^\circ$. $\langle u^2 \rangle$ was used for the Debye-Waller factor in eq.~(\ref{eq:Debye-Waller}). The three right columns compare our obtained parameters with the literature.\newline 
   $^{(*)}$ Note that each unit cell contains two layers of alanin-strands in $x$-direction. Therefore $M_x = 1.5$ corresponds to three layers of $\beta$-sheets in a single crystallite.\newline
   $^{(**)}$ Statistical model: A layer is shifted by a value $+a_y/4$ or $-a_y/4$ with respect to the previous layer, where $+$ and $-$ are equally likely.}
 \label{tab:parameters}
\end{table*}

From Eq.~(\ref{eq:ScatteringAmplitude}) it is clear that the
$z$-components of the atom positions $\{{\bf r}_k\}_{k=1}^K$ are
irrelevant for the scattering amplitude $A({\bf q})$ in the $xy$-plane, i.e.~if the
$z$-component of ${\bf q}$ is zero.  Therefore, parameters affecting
only the $z$-components -- especially the mentioned shifts in the
$z$-direction -- will not influence the intensity profile of $G({\bf
  q})$ in the $xy$-plane.\footnote{In principle there can be an
  influence because of the $\theta$-tilt of the crystallites with
  respect to the fiber axis (see section \ref{sec:Ensemble} and
  fig.~\ref{fig:EulerAngles}). However, the scattering amplitude
  $A({\bf q})$ shows a \emph{discrete} peak structure and, for small
  $\theta$-rotations, the out-of-plane reflections (which \emph{are}
  influenced by the $z$-components) are too far away to have an impact
  on the in-plane intensity
  profile.}
Analogously, the scattering profile in the $z$-direction is
independent of parameters influencing the $x$- and $y$-directions.
Consequently, the sections of the scattering profile along and
perpendicular to the fiber axis can be matched to subsets of the
parameters separately. The intensity profile off the $z$- and
$xy$-axes, taking into account all dimensions of the crystallite, can
be seen as a consistency check for the found parameters.

\begin{figure*}[h]
 \begin{minipage}[t]{\textwidth}
   \includegraphics[width=6cm]{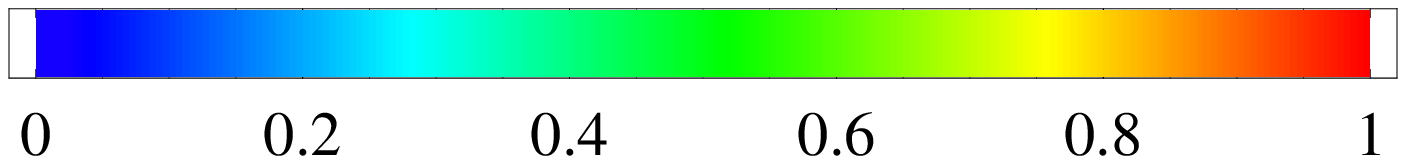}
   \centering
 \end{minipage}

 \begin{minipage}[t]{0.49\textwidth}
   \includegraphics[height=7cm]{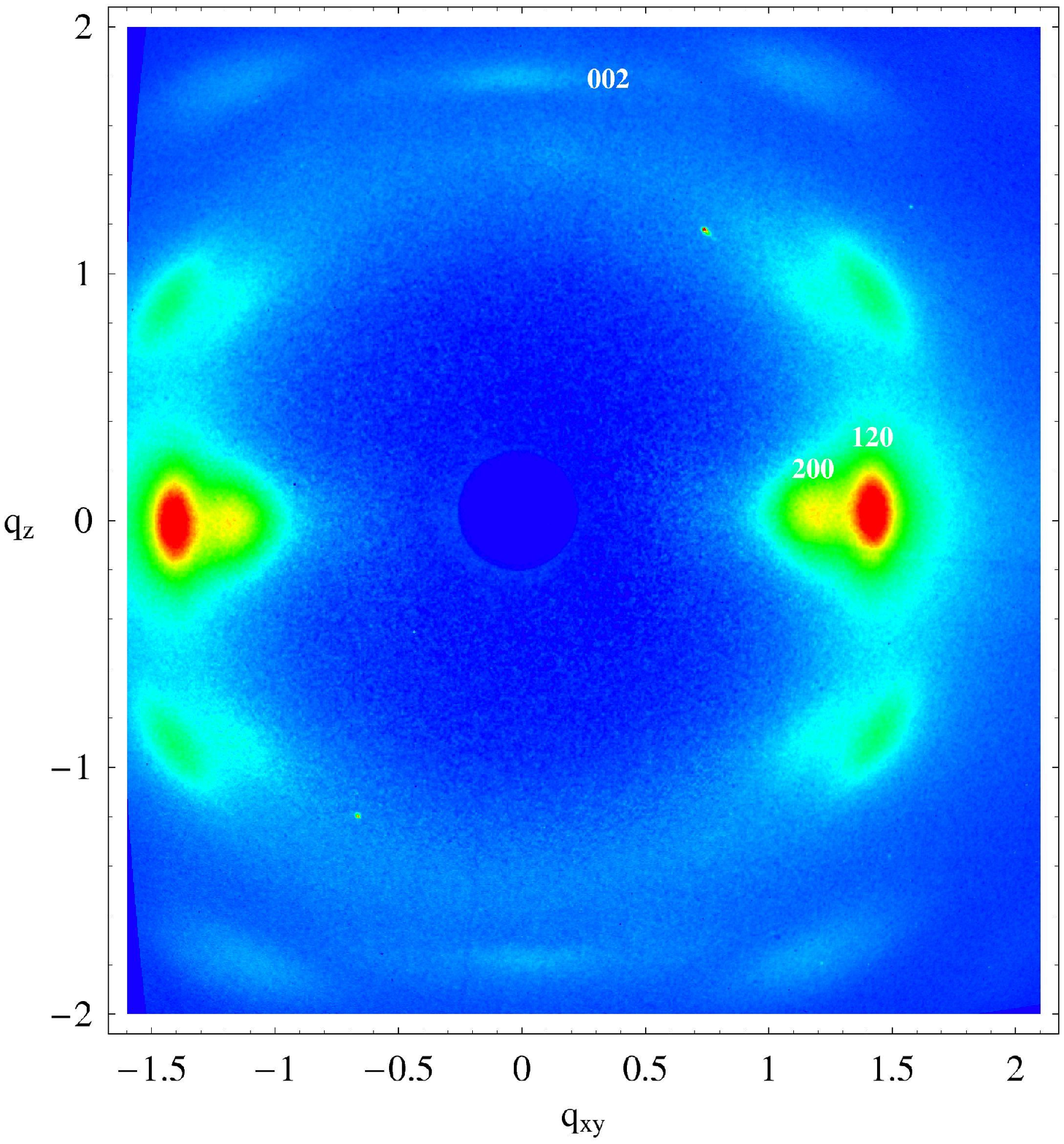}
   \centering
 \end{minipage}
 \begin{minipage}[t]{0.49\textwidth}
   \includegraphics[height=7cm]{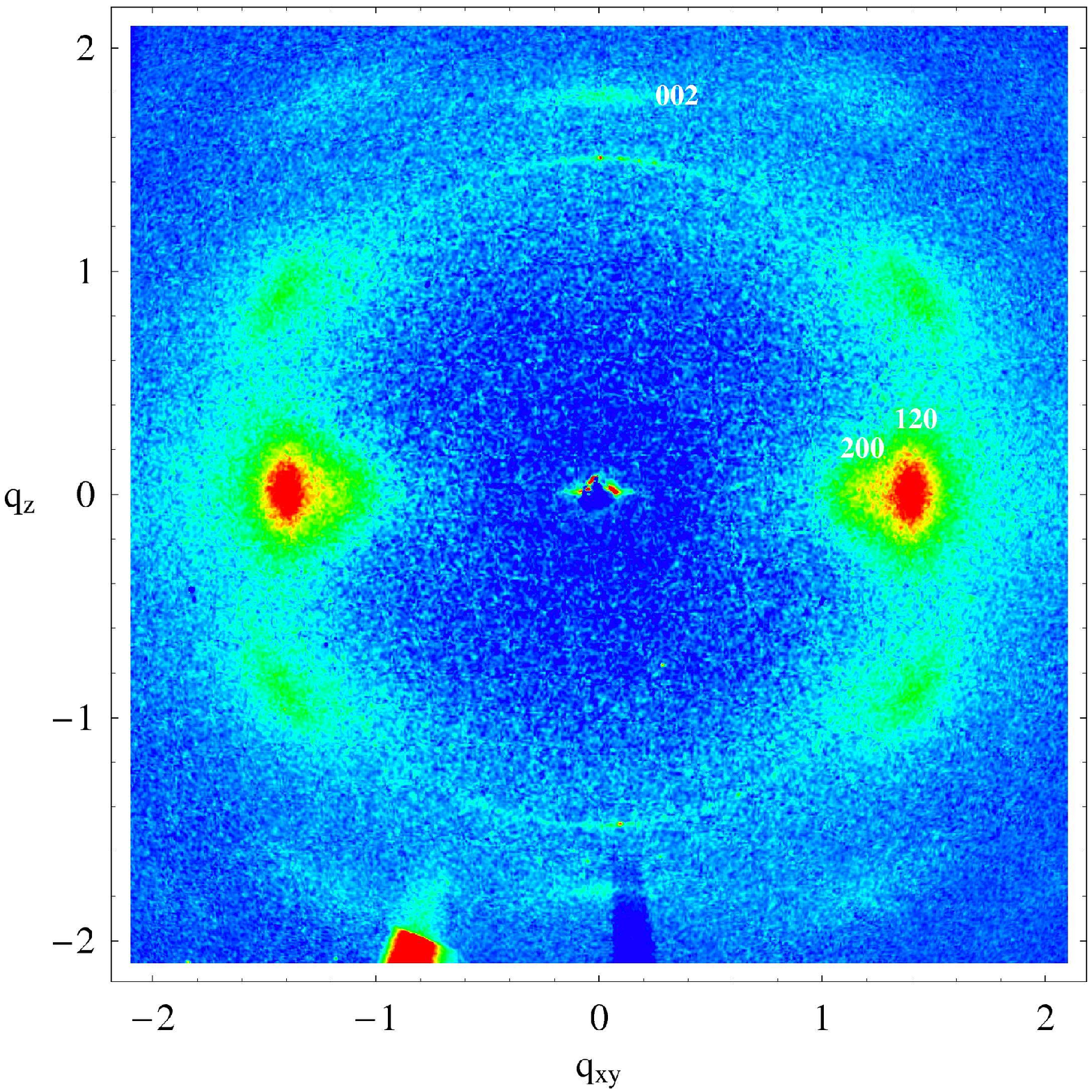}
   \centering
 \end{minipage}
\caption{(Color online) Scattering images of spider silk. The fiber
   axis runs vertical. On top, the colorbar shows scattering
   intensities, which are normalized by the intensity of the
   (120)-peak. Below the experimental scattering image of
   spider silk from \emph{Nephila clavipes} is shown, both for bundle 
   measurements (left) and single fiber diffraction (right)}
\label{fig:experimentsalditt}
\end{figure*}
\begin{figure*}[h]
 \begin{minipage}[t]{.49\textwidth}
   \includegraphics[height=7cm]{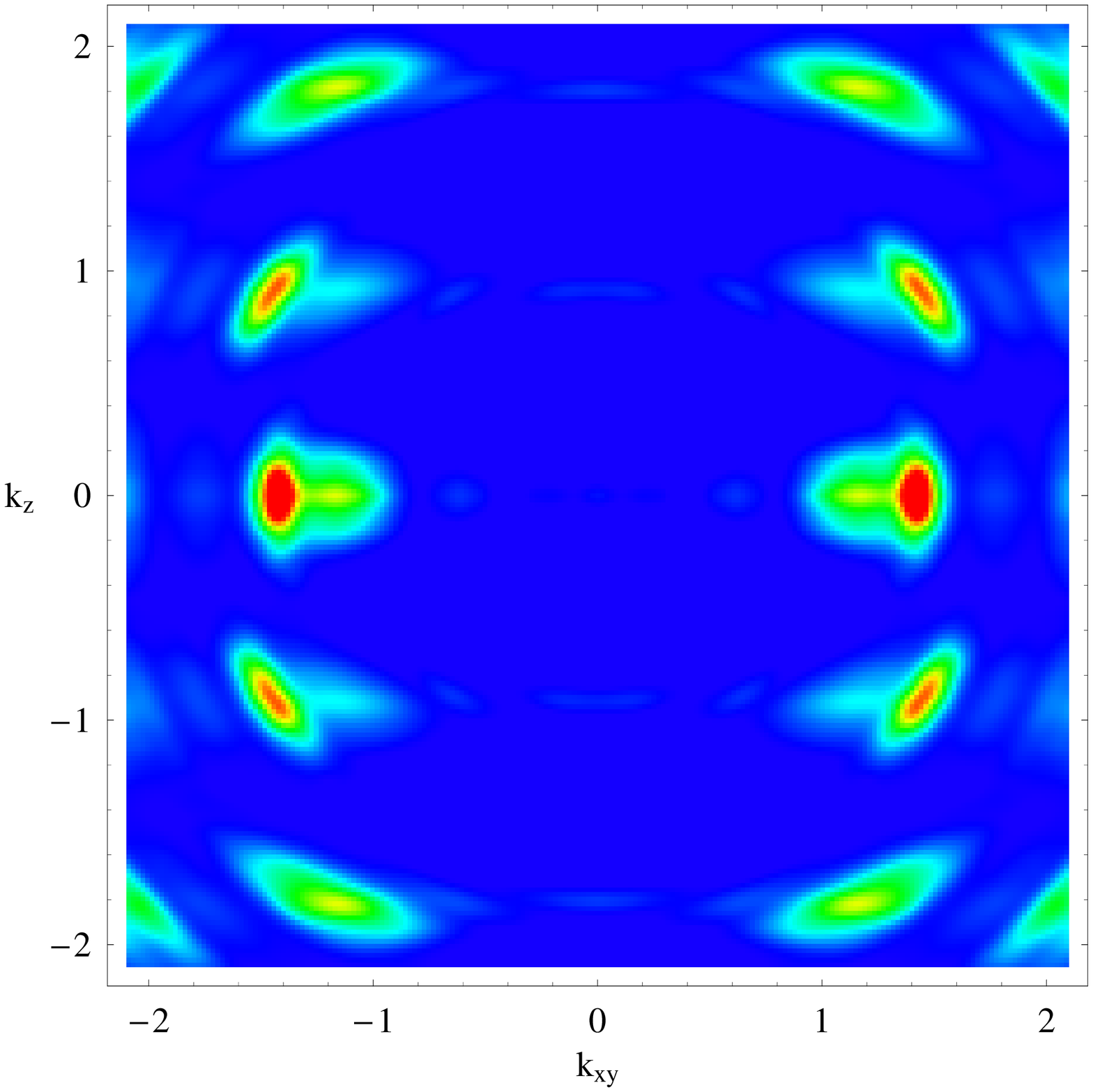}
   \centering
 \end{minipage}
\begin{minipage}[t]{.49\textwidth}
   \includegraphics[height=7cm]{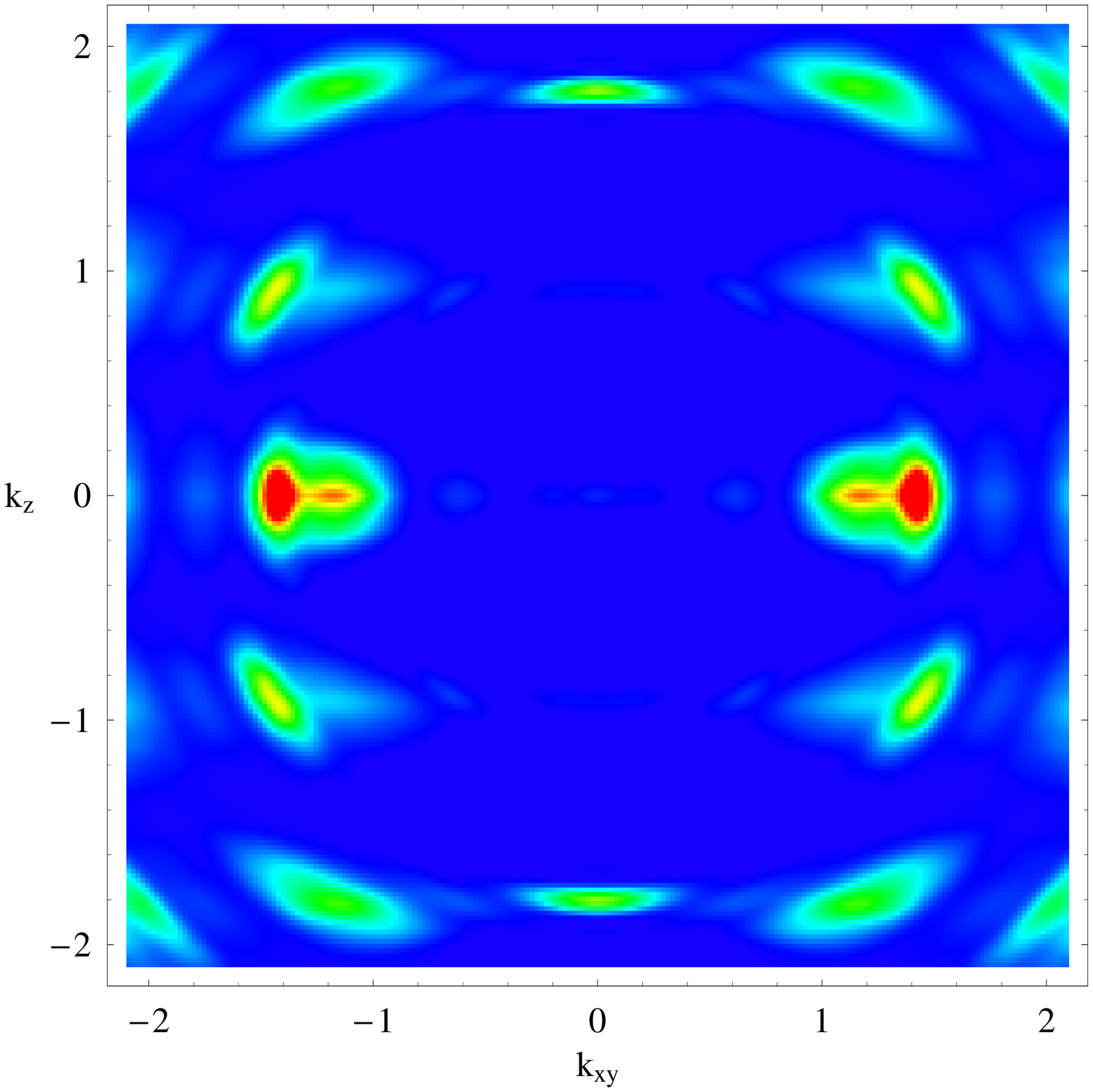}
   \centering
 \end{minipage}

 \caption{(Color online) Calculated scattering images of the structure
   proposed by Marsh \emph{et al.} \cite{marshShort}. Unit cell size:
   $(a_x,a_y,a_z) = (10.6,9.44,6.95)$ and we used the crystallite size
   $(M_x,M_y,M_z)=(1.5,6,9)$. Left: The crystallites are purely made
   of alanine amino acids. Right: The crystallites' alanine amino
   acids are replaced with glycine with a probability
   $p_\text{gl}=0.375$.}
 \label{fig:ScatteringImageSizeMarsh}  
\end{figure*}

The experimental scattering data clearly reveal a (002) peak, 
Fig.~\ref{fig:experimentsalditt}. This peak is allowed by symmetry, however it
is extremely weak in the antiparallel structure suggested by Marsh \cite{marshShort} and
shown in Fig.~\ref{fig:ScatteringImageSizeMarsh}. The reason is the
following: the electron density within the unit cell projected along the $z$-axis is almost
uniform, varying by approximately $10\%$. We therefore propose two alternative
mechanisms generalizing the classical (Marsh) model of the antiparallel unit cell.
By both mechanisms  the intensity of the (002) peak will increase in agreement with the experiment:
\begin{itemize}
\item[a)] the shift of strands 2 and 4 in the $z$-direction, i.e. a nonzero
$\Delta z_{24}$-shift or
\item[b)] structural disorder affecting the almost uniform electron density.
\end{itemize}

We first discuss case a). The uniform electron density is disturbed by
a shift $\Delta z_{24}\neq 0$. The intensity of the (002)-reflection grows accordingly
with an increasing shift $\Delta z_{24}$. Adjusting the $\Delta
z_{24}$-shift yields results consistent with experiment.

In table \ref{tab:parameters} we present the results for the
parameters of the model, obtained from optimising the agreement
between the calculated scattering function and the experimental one.
For comparison we show the set of parameters for \emph{both}, the parallel
and the antiparallel structure. On the basis of the experimental
data, one can not discriminate between the parallel and the antiparallel
structure. 

\begin{figure*}[h]
 \begin{minipage}[t]{.49\textwidth}
   \includegraphics[width=7cm]{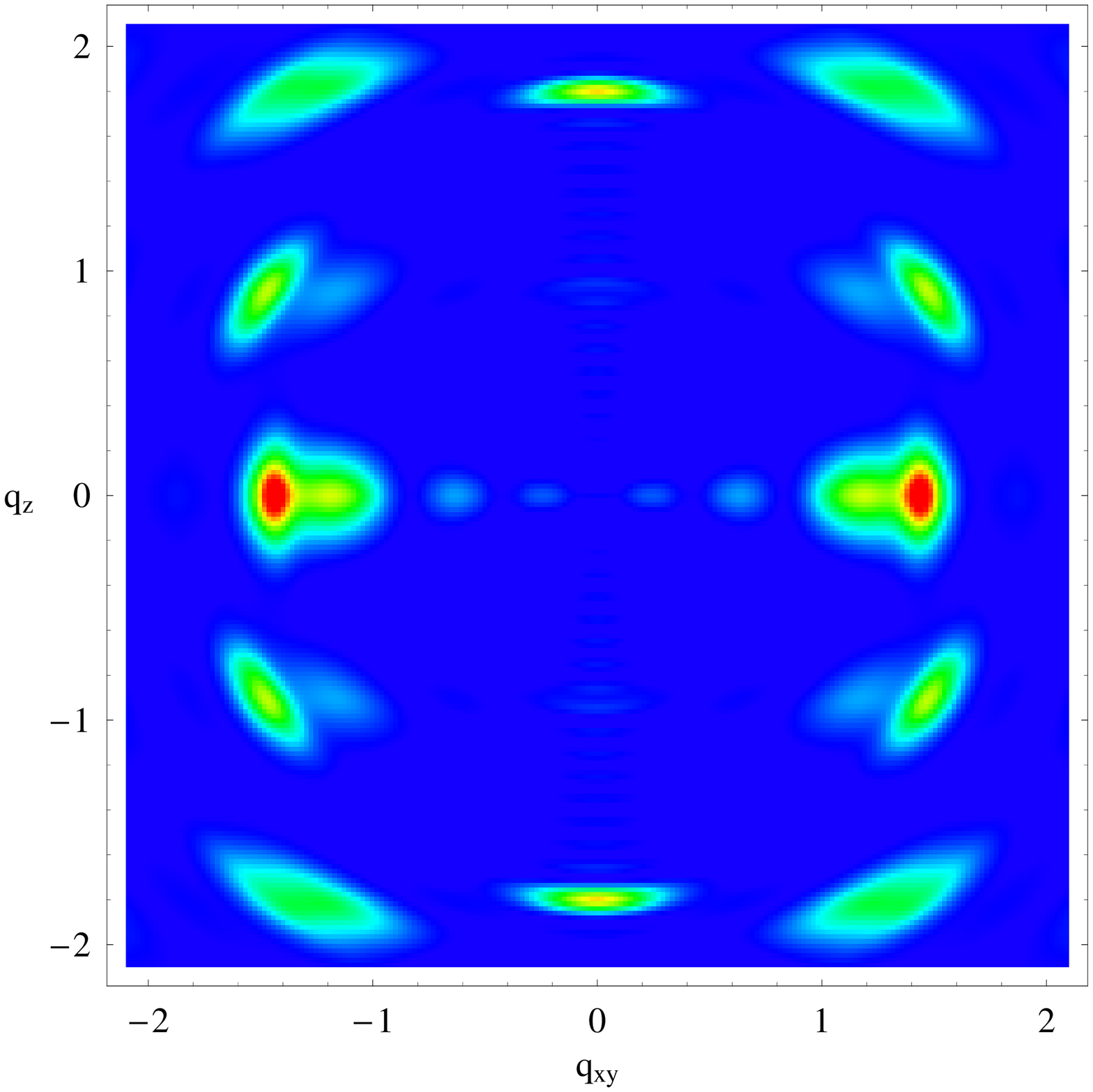}
   \centering
 \end{minipage}
\begin{minipage}[t]{.49\textwidth}
   \includegraphics[width=7cm]{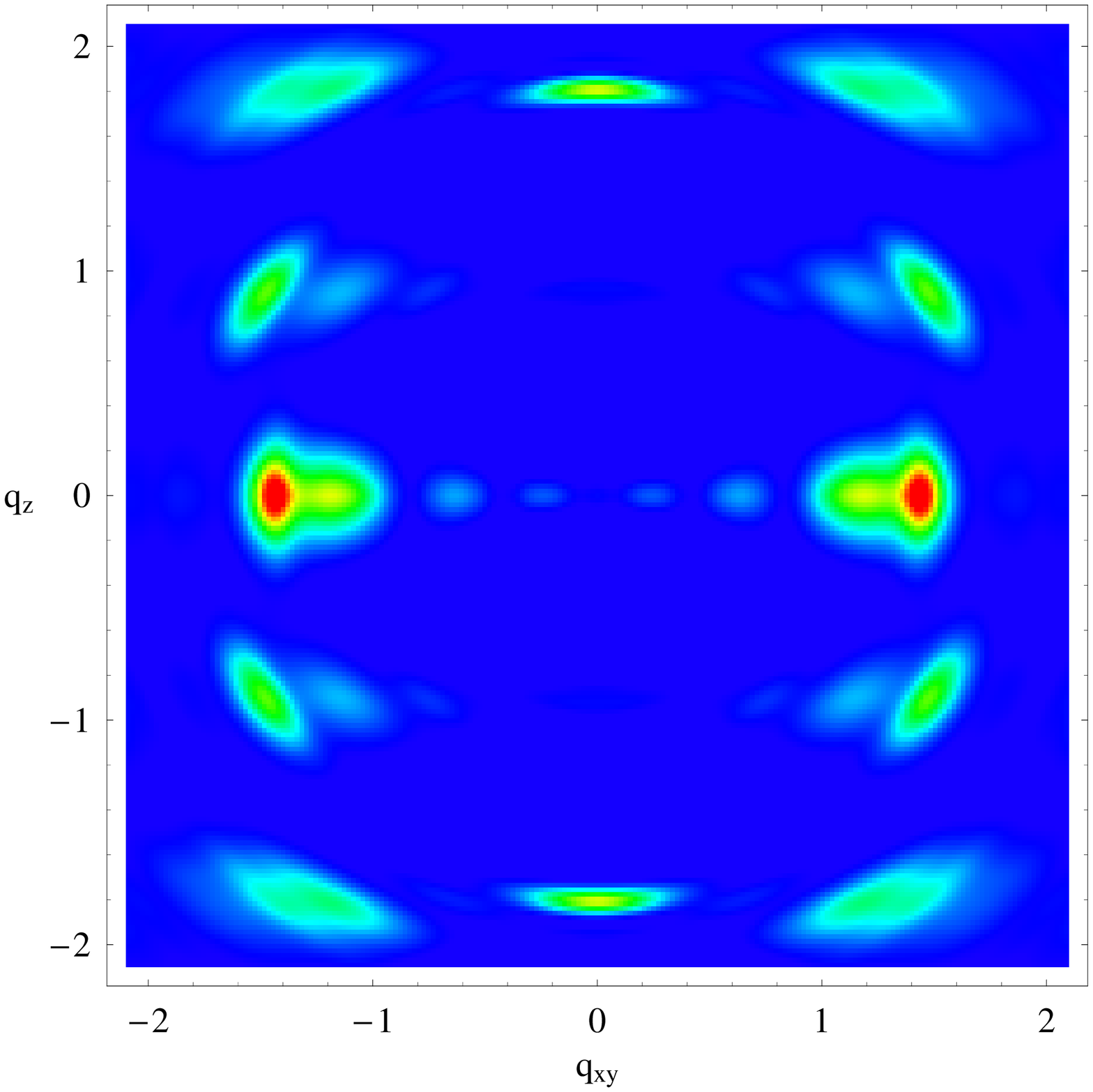}
   \centering
 \end{minipage}

 \caption{(Color online)  Scattering images, as calculated from
   eq.~(\ref{eq:ScatteringFunctionFinal}), for the parallel structure
   on the left side and the antiparallel structure on the right side.}
 \label{fig:ScatteringImage}  
\end{figure*}

\begin{figure}[h]
\centering
\begin{minipage}{.4\textwidth}
   \includegraphics[width=7cm]{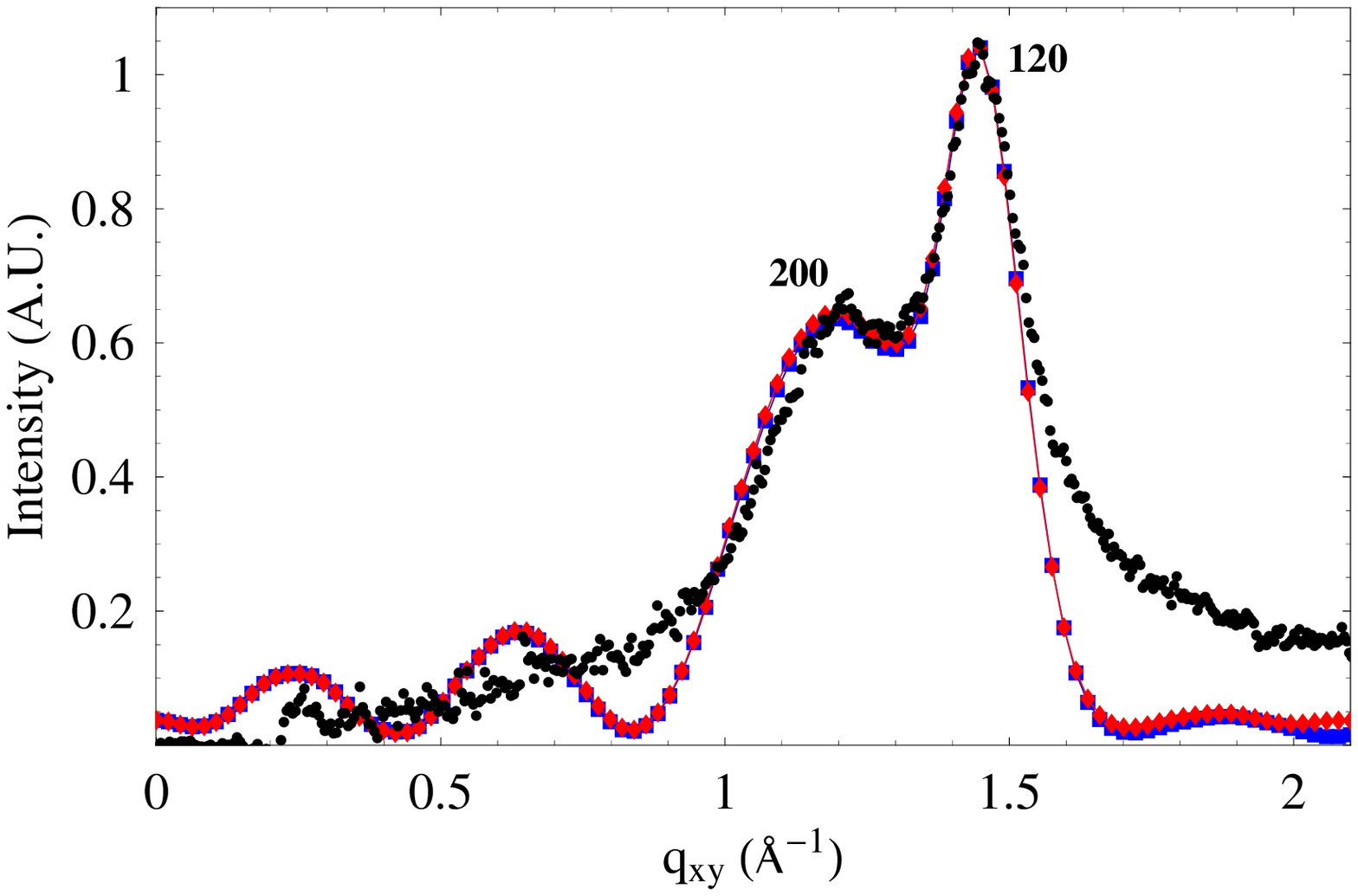}
   \centering
\end{minipage}

\begin{minipage}{.4\textwidth}
   \includegraphics[width=7cm]{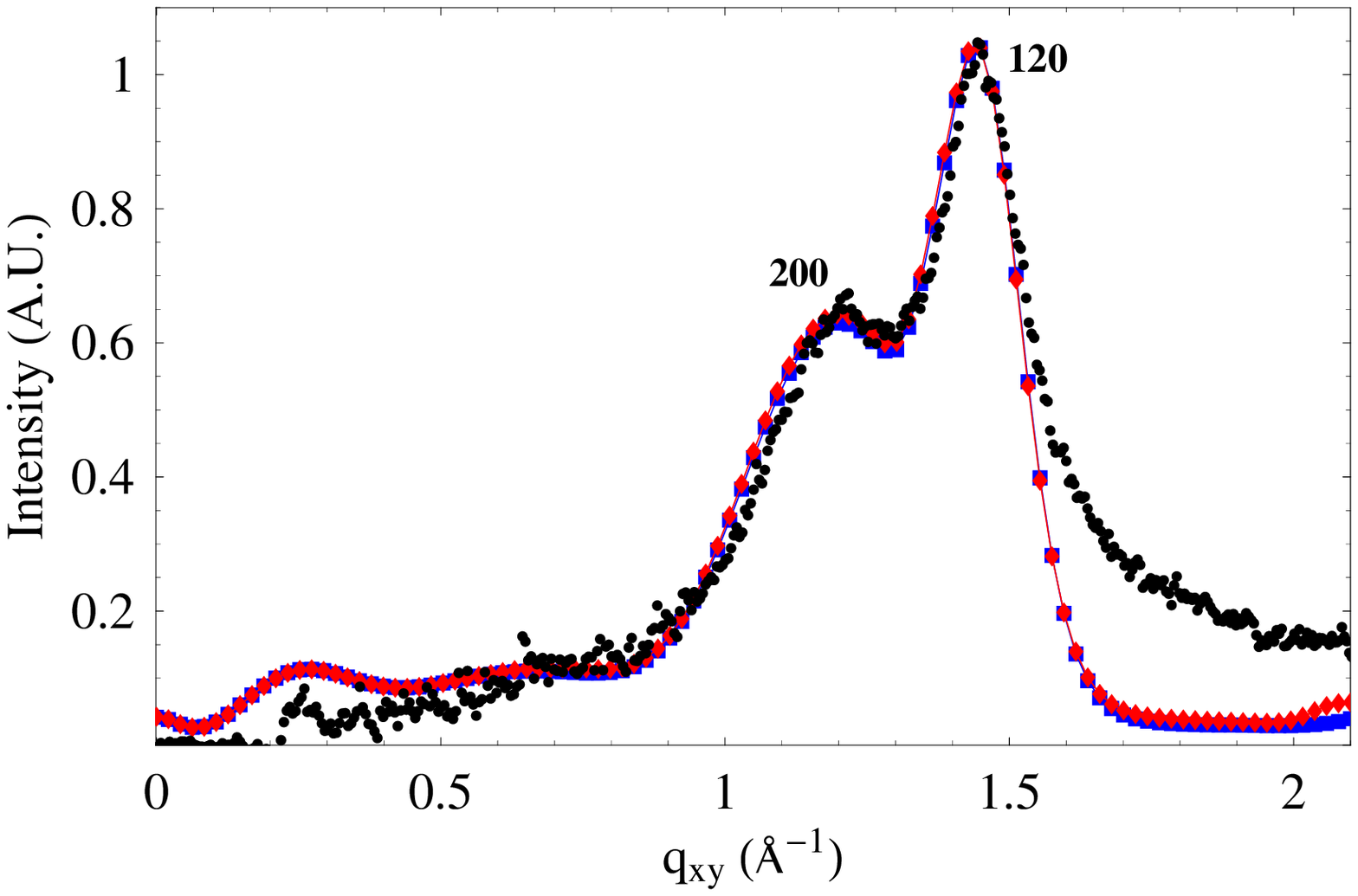}
   \centering
\end{minipage}
\caption{(Color online) Top: Comparison of a section of the experimental ($\bullet$) and the calculated scattering intensity 
( $\textcolor{blue}{{\scriptstyle\blacksquare}\!\!\!\!\!\!\frac{\;\;\;\;\,}{\,}}$ parallel,~~$\textcolor{red}{{\blacklozenge}\!\!\!\!\!\!\frac{\quad}{\,}}$ antiparallel). Sections of the profiles in Fig.~\ref{fig:ScatteringImage} along the $q_{xy}$-axis, i.e.~the scattering profile perpendicular to the fiber axis, are shown.
Bottom: as top, but with a Gaussian distribution (rounded to integers) of the crystallite sizes $M_x$, $M_y$ and $M_z$. The widths are $\Delta M_x = 2$, $\Delta M_y = 0.75$ and $\Delta M_z = 3$ respectively.}
 \label{fig:ScatteringImage:1d}  
\end{figure}

The scattering intensities, as calculated with these values, are
shown in Fig.~\ref{fig:ScatteringImage}.
The crystallites are randomly tilted with respect to the fiber axis,
so that on average the system is invariant under rotations around the
fiber axis. Consequently the scattering image also has rotational
symmetry about the $z$-axis and the $q_x$- and $q_y$-axis are
indistinguishable and denoted by $q_{xy}$. A section along the
$q_{xy}$-axis is shown in fig.~\ref{fig:ScatteringImage:1d}, top panel. The
mismatch for $q$-values slightly larger than the (120)-peak is
plausible, because in this region the amorphous matrix contributes
noticeably to the experimental scattering intensity, but has been
neglected in the model. The oscillations of the calculated scattering
image for low $q$-values are side maxima which are suppressed by
fluctuations in the crystallite sizes (see sec.~\ref{sec:crystVariations}). 
The corresponding scattering intensities 
are shown in Fig.~\ref{fig:ScatteringImage:1d} and Fig.~\ref{fig:ScatteringImageCorrelated}, left. Clearly the side maxima
have been suppressed.

We now discuss an alternative mechanism to generate a stronger (002) peak, e.g. by
introduction of disorder in the amino acid composition of the unit cell (case b).  
Poly-alanine as a model for the crystallites in spider silk is an over-simplification, since the 
amino acid sequence hardly allows for a pure poly-alanine crystallite. Instead we expect that other 
residues must be incorporated into the crystallite even if energetically less favorable to compromise
the given sequence. In particular, it is highly likely that also glycine amino acids are embedded in the
crystallites \cite{marsh}. This can be easily
implemented by replacing randomly selected alanine amino acids of the
crystallites with glycine (see sec.~\ref{sec:crystVariations}).  
It is found that the intensity of the (002) peak increases with the fraction of
susbstituted alanines. In Fig.~\ref{fig:ScatteringImageSizeMarsh} we
compare the original Marsh-structure (without gylcine) to a structure
with the same parameters, but with alanine replaced with glycine
randomly with a probability $p_\text{gl} = 0.375$. The random
substitution has clearly produced an intensity of the (002) peak
comparable to experiment.

\section{Conclusions}

We have developed a microscopic model of the structure of spider
silk. The main ingredients of the model are the following:

\begin{itemize}
\item[a)] Many small crystallites are distributed randomly in an amorphous
matrix,

\item[b)] the orientation of the crystallites fluctuate with a preferential alignment along the
fiber axis,

\item[c)] each crystallite is composed typically of $5\times 2\times 9$ unit
cells,

\item[d)] each unit cell contains four alanine strands, constructed with
Yasara and shifted with respect to each other. Disorder can be generated by randomly replacing alanine with glycine.

\end{itemize}

We have computed the scattering intensity of our model and
compared it to wide-angle x-ray scattering data of spider silk.
Possible inter-crystallite correlations are unimportant, given the
measured orientational distribution.  In other words, even if
significant center-of-mass correlations between crystallites were
present, the orientational distribution would suppress interference
effects, with the exception of the (002) peak, which is least
sensitive to orientational disorder. The contribution of coherent scattering is discussed in detail in Appendix~\ref{sec:irrelevantCoherent}.

A homogeneous electron density background is a necessary feature of
the scattering model. Calculation of the crystal structure factor in
vacuum does not only lead to an incorrect overall scaling prefactor
(which is important if absolute scattering intensities are measured),
but also leads to a scattering intensity distribution with artifacts
at small and intermediate momentum transfer.

The comparison between model and data fixes the parameters of the unit
cell and the crystallite for the two possible cases, the parallel and
the antiparallel structure, respectively, as shown in
Table~\ref{tab:parameters}. The two models with parallel and antiparallel alignment
of the alanine strands yield comparable agreement with the
experimental data. Also a refined model in which alanine is randomly replaced with glycine give reasonable results. Hence we cannot rule out one of these
structures.

Our model is similar to the model of the poly-L-alanine of Arnott \emph{et al.} \cite{arnott67}. Their model does incorporate a $\Delta z_{24}$-shift. However, our structure shows a better agreement with the experimentally measured scattering function using a value of $\Delta z_{24} = -a_z/6$.

While we have concentrated here on the wide-angle scattering
reflecting the crystalline structure on the molecular scale, we note
that the same model can be used for small-angle scattering to analyze
the short range order between crystallites in the presence of
orientational and positional fluctuations. In particular, the model
can describe the entire range of momentum transfer and the transition
from wide angle scattering (WAXS) to small angle scattering
(SAXS). Note that WAXS is usually described only in the single
object approximation, neglecting inter-particle
correlations. Contrarily, SAXS is mostly described in contiunuum
models without crystalline parameters. Here both are treated by the
same approach, which is a significant advantage for systems where the
length scales are not decoupled.

\subsubsection*{Acknowledgments}
We thank Martin Meling for a beneficial discussions and cooperation on the construction of the parallel and antiparallel raw structures of the $\beta$-sheets. We gratefully acknowledge financial support by the Deutsche Forschungsgemeinschaft (DFG) through Grant SFB 602/B6.

\section*{Appendix}
\appendix

\section{Effect of the continuous background\label{sec:appBackground}}
In this section we explain why it is necessary to include the continuous background between the crystallites, introduced in section \ref{sec:background}. 

\begin{figure}[h]
   \includegraphics[width=.45\textwidth]{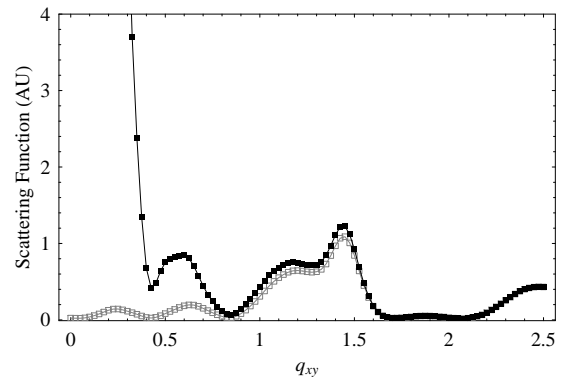}
   \centering
 \caption{Scattering function in $xy$-direction with ($\scriptstyle\square$) and without ($\scriptstyle\blacksquare$) continuous background. Without background, density fluctuations on large length scales cause an increase of the scattering function for small $q$-values.}
  \label{fig:PlotBackground}
\end{figure}

Without the background, the system has vast, unphysical density fluctuations on the length scale of the crystallite distances, resulting in a large scattering function $G({\bf q})$ for small $q$-values. As already explained in section \ref{sec:background}, these density fluctuations are unphysical, because the space between the crystallites is filled with the amorphous matrix and water molecules. Fig.~\ref{fig:PlotBackground} shows the the scattering profiles in $xy$-direction with and without the continuous background. As expected the system without the background shows a large increase of the scattering function for small $q$-values. The countinous background, howerver, acts as a low-pass filter on the scattering density and therefore annihilates the large intensities for small $q$.

\section{Relevance of the coherent part of the scattering function} \label{sec:irrelevantCoherent}
\begin{figure*}[h]
 \begin{minipage}[t]{.99\textwidth}
   \includegraphics[width=6cm]{fig/Hue2Vert1.eps}
   \centering
 \end{minipage}

\begin{minipage}{.49\textwidth}
   \includegraphics[height=7cm]{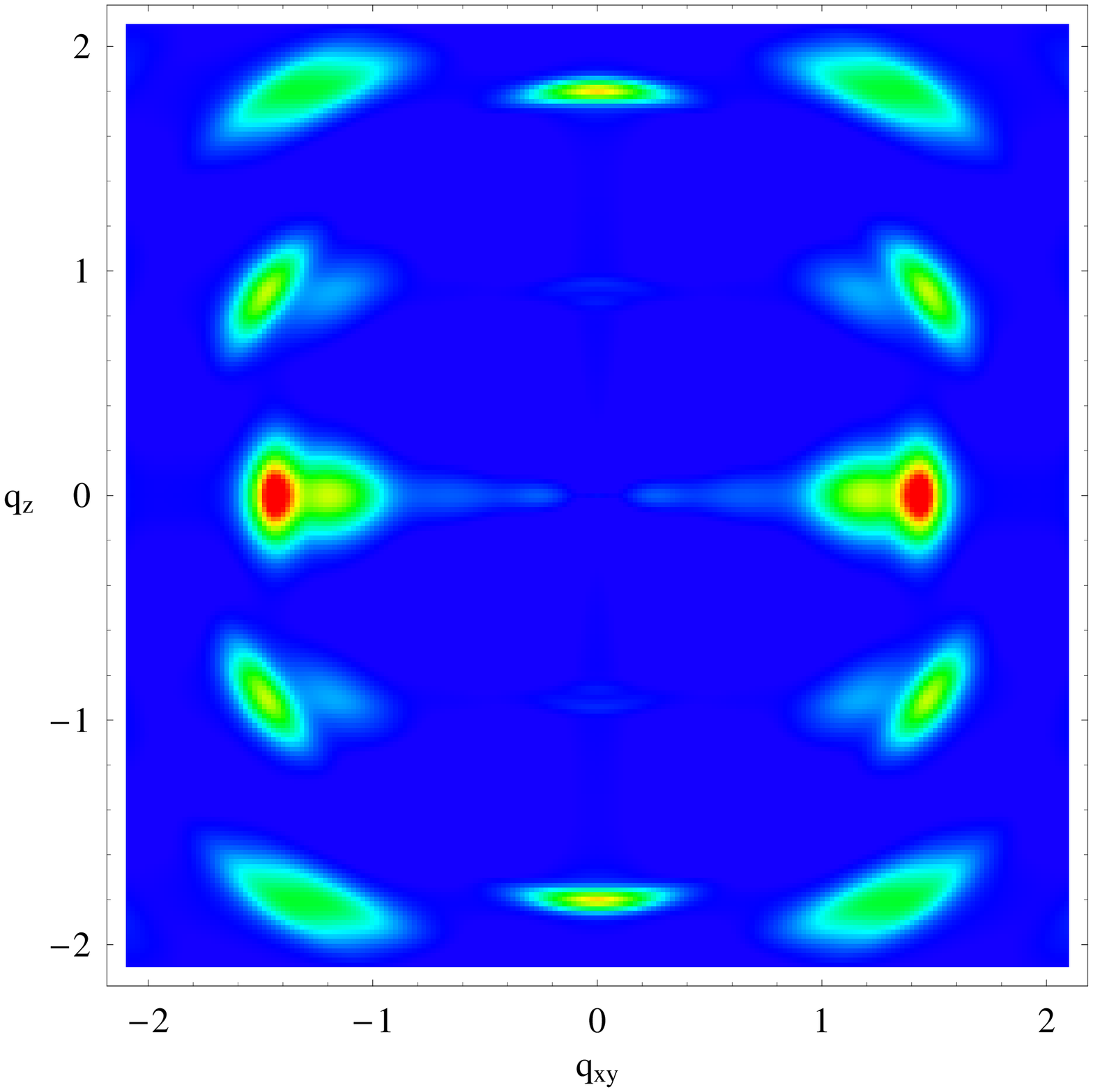}
   \centering
\end{minipage}
\begin{minipage}{.49\textwidth}
   \includegraphics[height=7cm]{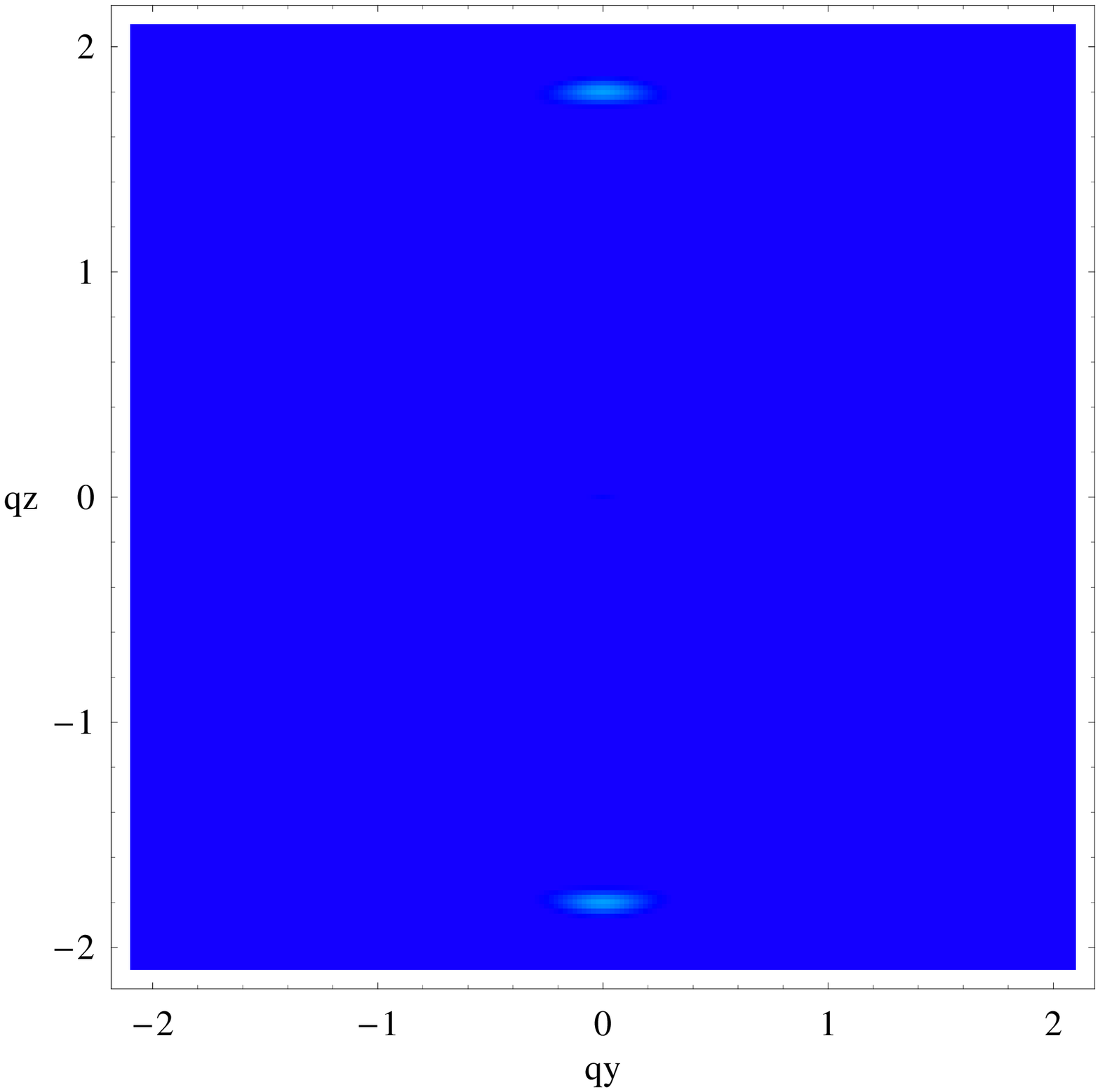}
   \centering
\end{minipage}

\caption{(Color online) Left: Calculated scattering images as in
  Fig.~\ref{fig:ScatteringImage}, but for a Gaussian distribution
  (rounded to integers) of the crystallite sizes $M_x$, $M_y$ and
  $M_z$. The widths are $\Delta M_x = 2$, $\Delta M_y = 0.75$ and
  $\Delta M_z = 3$ respectively.  Right: \emph{correlated} part
  $G'_2({\bf q})$ of the scattering function
  eq.~(\ref{eq:ScatteringFunctionFinal}) in comparison to the
 \emph{uncorrelated} part (left).} 
 \label{fig:ScatteringImageCorrelated}  
\end{figure*}

Here we discuss the infulence of the coherent part of the scattering function of eq.~(\ref{eq:ScatteringFunctionFinal}). Fig.~\ref{fig:ScatteringImageCorrelated} shows a comparison of the incoherent part 
\begin{equation}
G_1({\bf q}) = \int \mathcal{D}\underline{\underline{D}} \left| A(\underline{\underline{D}}^T {\bf q})  \right|^2 \,, 
\end{equation}
which is used to calculated the scattering function in this paper, and the contribution 
\begin{equation}
G'_2({\bf q}) := \left|  \int \mathcal{D}\underline{\underline{D}} \, A(\underline{\underline{D}}^T {\bf q})  \right|^2 
\end{equation}
of the coherent part $G_2({\bf q}) = \left(S({\bf q}) - 1 \right) \left|  \int \mathcal{D}\underline{\underline{D}} \, A(\underline{\underline{D}}^T {\bf q})  \right|^2$. 

Neglecting the coherent part is plausible for two reasons. Firstly, because the contribution of $G'_2({\bf q})$ is small compared to the incoherent part $G_1({\bf q})$, as seen in the figure. And secondly, the length scale for the distances between the crystallites is much larger than atom length scales investigated here. 
On length scales we are interested in, we expect $S({\bf q}) \approx 1$, assuming that the crystallite positions have no long range order. Therefore, the prefactor $\left(S({\bf q}) - 1 \right)$ additionally reduces contribution of the coherent term.

The (002)-peak is special for the coherent scattering term $G'_2({\bf q})$. All peaks except for the (002)-peak have a very small contribution in $G'_2({\bf q})$ because of the white average of the crystallites' rotations about the fiber axis, which makes coherent scattering from \emph{different} crystallites less likely, no matter how the crystallites are arranged in space. 
Since there is a preferential alignment of the crystallites in $z$-direction, however, contributions of coherent scattering from different crystallites (which are contained in the term $G'_2({\bf q})$) are not completely destroyed; therefore, if the crystallites' distance in $z$ direction is a multiple of the unit cell size $a_z$, causing a large contribution in the prefactor $\left(S({\bf q}) - 1 \right)$ at the position of the (002)-peaks, a contribution of $G_2({\bf q})$ would be present.

\bibliographystyle{epj}
\bibliography{SpiderSilk}


\end{document}